%% file: arxiv_v2.tex
\documentclass[aps,preprintnumbers,superscriptaddress,nofootinbib,aps,prd,twocolumn,floatfix]{revtex4-1}
\pdfoutput=1

\RequirePackage[colorlinks=true
,urlcolor=blue
,anchorcolor=blue
,citecolor=blue
,filecolor=blue
,linkcolor=blue
,menucolor=blue
,pagecolor=blue
,linktocpage=true
,pdfproducer=medialab
,pdfa=true
]{hyperref}

\usepackage{amsmath,amssymb}
\usepackage{graphicx}
\usepackage{url}
\usepackage{slashed}
\usepackage{multirow}
\usepackage{footmisc}
\usepackage{hyperref}
\usepackage{braket}
\usepackage{colortbl}
\usepackage{booktabs}

\graphicspath{{./figs/}}

\input{declare}

\begin{document}
\title{Higgs Pair Production at Future Hadron Colliders: From Kinematics to Dynamics}

\author{Dorival Gon\c{c}alves} 
%\email{dorival@pitt.edu}
\affiliation{PITT PACC, Department of Physics and Astronomy, University of Pittsburgh, USA} 

\author{Tao Han}
% \email{than@pitt.edu}
\affiliation{PITT PACC, Department of Physics and Astronomy, University of Pittsburgh, USA} 

\author{Felix Kling}
\affiliation{Department of Physics and Astronomy, University of California, Irvine, USA}

\author{Tilman Plehn}
\affiliation{Institut f\"ur Theoretische Physik, Universit\"at Heidelberg, Germany}

\author{Michihisa Takeuchi} 
%\email{michihisa.takeuchi@ipmu.jp}
\affiliation{Kavli IPMU (WPI), UTIAS, University of Tokyo, Kashiwa, Japan}
 
\preprint{PITT-PACC-1802, UCI-TR-2018-1, IPMU-18-0028}

%%%%%%%%%%%%%%%%%%%%%%%%%%
\begin{abstract}
\noindent
The measurement of the triple Higgs coupling is a key benchmark for
the LHC and future colliders.  It directly probes the Higgs potential
and its fundamental properties in connection to new physics beyond the
Standard Model. There exist two phase space regions with an enhanced
sensitivity to the Higgs self-coupling, the Higgs pair production
threshold and an intermediate top pair threshold. We show how the
invariant mass distribution of the Higgs pair offers a systematic way
to extract the Higgs self-coupling, focusing on the leading channel
$pp\to hh+X\to b\bar b\ \gamma\gamma+X$. We utilize new features of the
signal events at higher energies and estimate the potential of a
high-energy upgrade of the LHC and a future hadron collider with
realistic simulations.  We find that the high-energy upgrade of the
LHC to 27 TeV would reach a 5$\sigma$ observation with an 
integrated luminosity of 2.5 ab$^{-1}$. It
would have the potential to reach 15\% (30\%) accuracy at 
the 68\% (95\%) confidence level to determine the SM Higgs boson self-coupling. A future 100 TeV collider could
improve the self-coupling measurement to better than 5\% (10\%) at the 68\% (95\%) confidence level. 
\end{abstract}

%%%%%%%%%%%%%%%%%%%%%%%%%%
\maketitle

\tableofcontents

%%%%%%%%%%%%%%%%%%%%%%%%%%
\section{Introduction}

The discovery of the Higgs boson~\cite{higgs,discovery} at the CERN
Large Hadron Collider (LHC) is of monumental significance. The
completion of the Standard Model (SM) provides us with a consistent
theory valid up to high scales. As a perturbative gauge theory, it
allows for precision predictions for essentially all LHC
observables. In parallel, experimental advances have turned ATLAS and
CMS into the first hadron collider precision experiments in
history. In combination, these developments open new avenues to tackle
fundamental physics questions at the LHC and future high-energy facilities.

On the theory side, we are still lacking an understanding of if and how
the Higgs mass, the only dimensionful parameter in the theory, is
stabilized against a large new physics scale. The Higgs potential
responsible for the electroweak symmetry breaking (EWSB) in the SM is
determined by the triple and quartic Higgs self-coupling
$\lambda_\text{SM}\approx 1/8$. It is a true self-interaction in the
sense that it is not associated with any conserved charge after
EWSB. With our ignorance for new physics beyond the SM, the shape of
the Higgs potential is deeply linked to the fundamental question of
electroweak symmetry breaking in the early universe, allowing for a slow second-order phase
transition in the SM or a strong first-order phase transition with a
modified Higgs potential.  It has been argued that a wide range of
modified Higgs potentials, which result in a strong first-order EW
phase transition, lead to order-one modifications of
$\lambda_\text{SM}$~\cite{ew_phase}.  All of this points to the Higgs
self-coupling $\lambda$ as a benchmark measurement for the coming LHC
runs, as well as any kind of planned colliders \cite{Arkani-Hamed:2015vfh}.
\medskip

Higgs pair production $pp\rightarrow hh$ offers a direct path to pin
down $\lambda$ at a hadron collider~\cite{hh-orig,hh-early}.  Previous
studies show that promising final states from the $hh$ decays are
$b\bar{b}\gamma\gamma$~\cite{hh-gamma,vernon},
$b\bar{b}\tau\tau$~\cite{hh-tautau-4b,hh-tautau},
$b\bar{b}WW$~\cite{hh-bbww}, $b\bar{b}b\bar{b}$~\cite{hh-4b}, and
$4W$~\cite{hh-ww}. Theoretical studies as well as current analyses
point to the $b\bar{b}\gamma\gamma$ decay as the most promising
signature at the LHC~\cite{current-gamma}. Combinations with indirect
measurements of the self-coupling from quantum effects confirm that
Higgs pair production provides the most robust self-coupling
measurement~\cite{indirect}.  For the high-luminosity LHC (HL-LHC),
ATLAS and CMS projections indicate a very modest sensitivity to the
Higgs self-coupling~\cite{hl-lhc}. 

In anticipation to probe new physics beyond the SM, it is customary to
parametrize the modification of the self-coupling as
\begin{align}
\kappa_\lambda = \frac{\lambda}{\lambda_\text{SM}} \; .
%= \kappa_\lambda \; \frac{3 m_h^2}{v} \; .
\end{align}
In the optimistic scenario that we can neglect systematic
uncertainties, those studies indicate that the LHC will probe the
coupling at 95\% confidence level
\begin{align}
-0.8 < \kappa_\lambda < 7.7\;.
\end{align}
An issue with those studies is that they are based on the total rate
for Higgs production, but neglect a wealth of available information.  Including
a full kinematic analysis could lead to an improved measurement
\cite{madmax-hh}
\begin{align}
-0.2< \kappa_\lambda <2.6 \;,
\end{align}
falling short in precision in comparison to other Higgs 
property measurements at the LHC, and far from satisfactory in probing the
Higgs potential.\medskip

In this study, we systematically compare the prospects for measuring
the Higgs self-coupling at current and higher energy $pp$
colliders. We focus on the two leading proposals for future hadron colliders: 
\begin{enumerate}
\item the 27~TeV high-energy LHC (HE-LHC) with an integrated
  luminosity of $15~\iab$,
\item a 100~TeV hadron collider with $30~\iab$, under consideration at
  CERN (FCC-hh)~\cite{europe_100tev} and in China
  (SppC)~\cite{china_100tev}.
\end{enumerate} 
We include state of the art signal and background estimates for the $b\bar{b} \gamma
\gamma$ channel, as well as realistic acceptance cuts and
efficiencies. While there exist a series of 100~TeV studies of Higgs
pair production at different levels of
sophistication~\cite{hh-nimatron}, we include a 100~TeV analysis to be
able to compare with the HE-LHC reach on equal footing.\medskip

We start with a study of relevant phase space regions using a
Neyman-Pearson maximum likelihood
approach~\cite{madmax-hh,madmax}. This allows us to estimate the
impact of using simple kinematic distributions on the measurement of
the Higgs self-coupling at the different colliders.  Furthermore, we
can evaluate the maximum significance of extracting the Higgs pair
signal and the significance of detecting a modified self-coupling
under idealized conditions.

In the main part of our paper, we perform a state-of-the-art analysis
of Higgs pair production including additional jet radiation and a full
set of realistic detector efficiencies. Unlike earlier analyses, we include 
$b$-jets from Higgs decays even when they become sub-leading in transverses momentum to the additional
jet radiation.  Our analysis focuses on the di-Higgs invariant mass
distribution, both for the extraction of the Higgs pair signal and for
the measurement of the Higgs self-coupling. Using a log-likelihood
approach on this single kinematic distribution, we show that the Higgs
self-coupling can be properly measured 
not only at a future 100~TeV
collider, but also at the 27~TeV HE-LHC.

%%%%%%%%%%%%%%%%%%%%%%%%%%
\section{Higgs Pair Signature}
\label{sec:frame}

\begin{figure}[t!]
  \includegraphics[width=.22\textwidth]{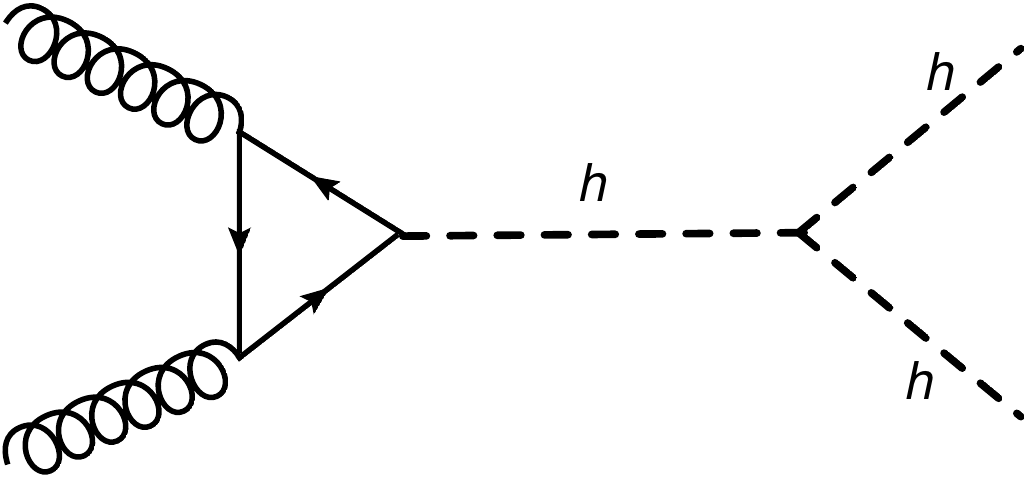}\hspace{0.2cm}
  \includegraphics[width=.22\textwidth]{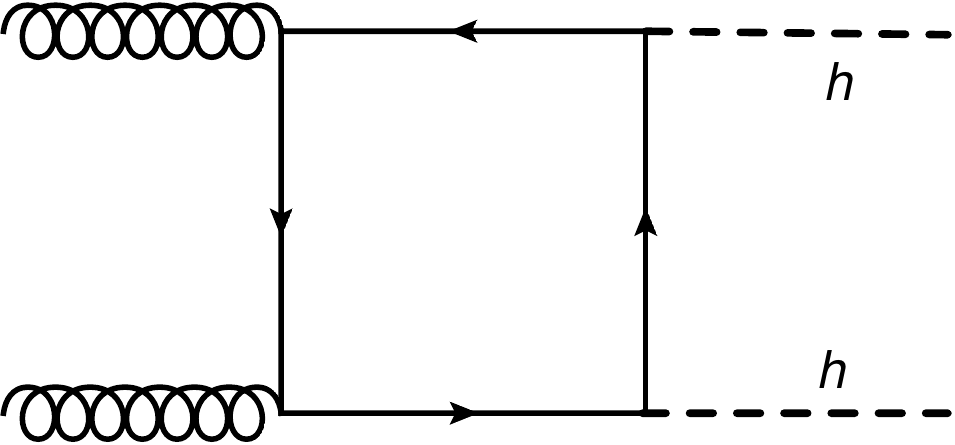}\\
 \caption{Representative Feynman diagrams contributing to the leading
   Higgs pair production process via gluon fusion.}
 \label{fig:feyn1}
\end{figure}

\begin{figure}[b!]
  \includegraphics[width=.44\textwidth]{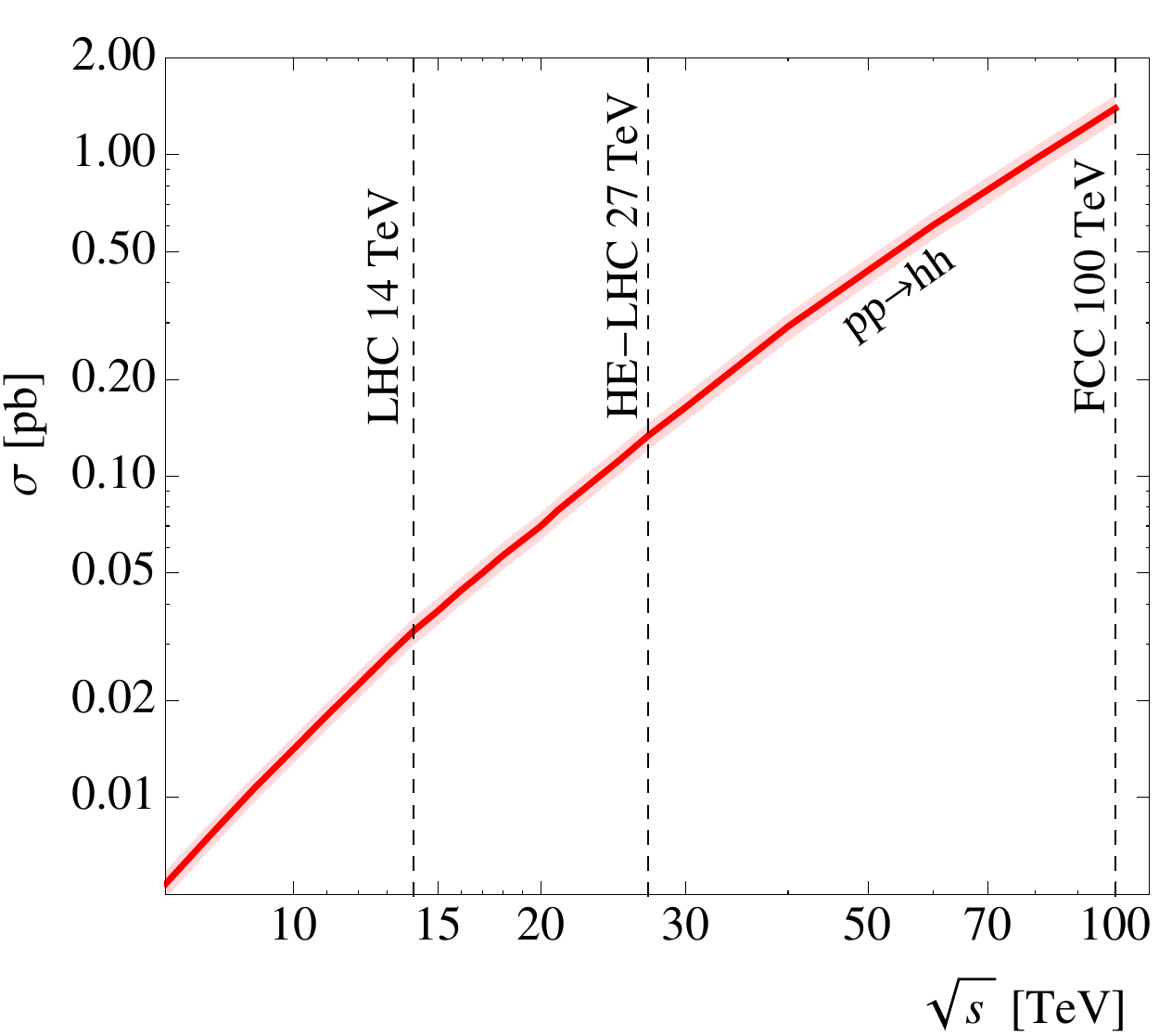}
 \caption{Total cross section for $pp\to hh$ production at NLO as a
   function of the $pp$ collider energy. The width of the curve
   reflects the 10\% theoretical uncertainty. }
 \label{fig:xs_hh_Ecm}
\end{figure}

The leading $hh$ production mechanism in the Standard Model at hadron
colliders is depicted by the Feynman diagrams in
Fig.~\ref{fig:feyn1}. Due to the difference of the top quark
propagators in the loops, the two diagrams interfere destructively.
In Fig.~\ref{fig:xs_hh_Ecm} we show the total rate for $hh$ production
as a function of the center of mass energy $\sqrt s$ in TeV, including
the next-to-leading order (NLO) corrections~\cite{nlo}.  The width of
the curve illustrated the theoretical uncertainties around
10\%~\cite{nnlo}.  At the LHC, the signal rate is the limiting factor
for Higgs pair studies.  At 14~TeV, the cross section including
higher-order corrections is in the range of 0.033~pb~\cite{nnlo},
corresponding to at most 100k events with an integrated luminosity of
$3~\iab$ at the HL-LHC. Assuming one Higgs decay to tagged bottom
quarks, the available rate is reduced to 60k events in the life time
of the HL-LHC. The crucial question is what kind of second Higgs decay
allows us to effectively trigger the events and to reduce the QCD
backgrounds to a manageable level. The leading candidate is the
signature~\cite{hh-gamma}
\begin{align}
pp \to hh \to b\bar{b} \; \gamma \gamma \; ,
\label{eq:bbgg}
\end{align}
because of the excellent di-photon mass resolution and the guaranteed
trigger. The expected number of signal events in the Standard Model at
the HL-LHC is 260. Alternatively, the $b\bar{b} \; \tau \tau$
signature leads to 7.2k events times the tau tagging probability rate
squared, and hampered by a significantly worse signal-to-background
ratio.\medskip

%-------------------------------------------------------
%-------------------------------------------------------

%-------------------------------------------------------
%-------------------------------------------------------

Because of the rapidly growing gluon luminosity at higher energies,
the $hh$ production cross section increases by about a factor of
4~(40) at 27~(100)~TeV.  This means that at the HE-LHC with the
anticipated integrated luminosity of $15~\iab$ the number of events in
the $b\bar{b} \; \gamma \gamma$ channel increases by a factor $4
\times 5 = 20$ to around 5k events.  A 100~TeV hadron collider with
a projected integrated luminosity of $30~\iab$ features another
increase by a factor $10 \times 2=20$, to around 100k expected Higgs
pair events in the Standard Model.

This estimate shows how the combination of increased energy and
increased luminosity slowly turns Higgs pair production into a valid
channel for precision measurements.  The numbers fundamentally affect
our proposed analysis strategy, because the small number of signal and
background events suggests a kinematic analysis including as few
kinematic distributions as possible.
It is possible to improve this situation, for example, using the matrix
element technique, as we will discuss below.\medskip

We
generate the signal with \textsc{MadGraph5}~\cite{mg5}, accounting for
a next-to-leading order (NLO) QCD factor $K_\text{NLO}\sim
1.6$~\cite{nlo}. In the final state we demand two $b$-tagged jets and
two isolated photons with the minimal acceptance and trigger cuts
\begin{alignat}{5}
&& p_{T,j}>30~\gev , \quad  |\eta_j |<2.5 \; , \notag \\
&& p_{T,\gamma}>30~\gev, \quad  |\eta_\gamma| <2.5 \; , \notag \\
&& \Delta R_{\gamma \gamma, \gamma j, jj} >0.4 \; .
\label{eq:base_selections}
\end{alignat}
The background to our $b\bar{b} \; \gamma \gamma$ signal consists of
other Higgs production modes ($t\bar{t}h, Zh$) with $h \to \gamma
\gamma$, continuum $b\bar{b}\gamma\gamma$ production, and of multi-jet
events with light-flavor jets faking either photons or $b$-jets
($jj\gamma\gamma, b\bar{b}\gamma j$)~\cite{hh-gamma}.  The different backgrounds  
are discussed in detail in Sec.~\ref{sec:ana}.

The proper simulation of efficiencies and fake rates are a key
ingredient for a realistic background estimate in this analysis.  For
the HE-LHC and the future 100~TeV collider we follow the ATLAS
projections~\cite{performance}. The efficiency for a tight photon 
identification can be well parametrized by
\begin{align}
\epsilon_{\gamma\to\gamma} = 0.863 - 1.07 \cdot e^{-p_{T,\gamma}/34.8~\gev}\;,
\end{align}
and a jet-to-photon mis-identification rate by
\begin{align}
\epsilon_{j\to\gamma} = 
\begin{cases} 
5.3\cdot 10^{-4} \exp \left( -6.5 \left( \dfrac{p_{T,j}}{60.4~\gev} - 1 \right)^2 \right)\;, \notag \\[4mm]
0.88 \cdot 10^{-4} \left[ \exp \left( -\dfrac{p_{T,j}}{943~\gev} \right) +\dfrac{248~\gev}{p_{T,j}}\right] 
\;,
\end{cases}
\end{align}
where the upper form applied to softer jets with ${p_{T,j} <65}$~GeV.
This leads to a photon efficiency of about 40\% at
$p_{T,\gamma}=30$~GeV, saturating around 85\% for
$p_{T,\gamma}>150$~GeV. Note that the Higgs decay products tend to be
soft, $p_{T,\gamma}\sim m_h/2$. 

For $b$-tagging, we adopt an efficiency with
\begin{align}
\epsilon_b =0.7 \; , 
\end{align}
associated with mis-tag rates of 15\% for charm quarks and 0.3\% for
light flavors.  These flat rates present a conservative estimate from
the two dimensional distribution on $(p_{Tj},\eta_j)$ shown in the
HL-LHC projections~\cite{madmax-hh}. Encouragingly, the small light
flavor fake rate projections result in a strong suppression for the
initially dominant $jj\gamma\gamma$ background.

Obviously, the final outcome of the analyses would depend on the detector performance for the efficiencies of photon identification and $b$-tagging, as well as the background jet rejection. To have a comprehensive exploration and comparison, we will also examine the other available detector parameters, one from CMS \cite{Chatrchyan:2012jua} and the other from the CERN Yellow Report \cite{Mangano:2017tke} for the future collider (FCC), as shown in the Appendix. 

%---------------------------------------------
\begin{figure*}[t]
  \includegraphics[width=0.32\textwidth]{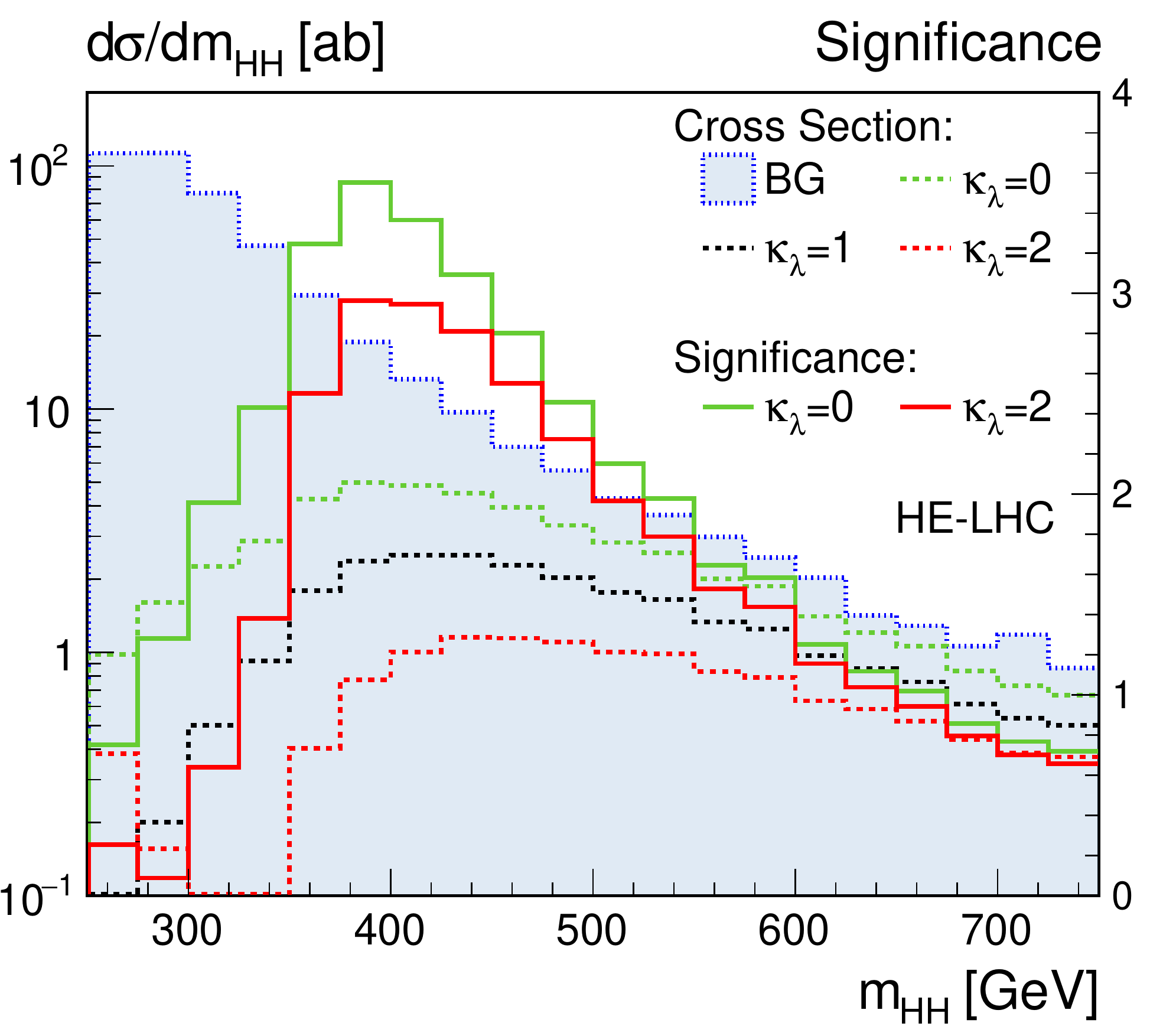}
  %\hspace*{0.02\textwidth}
  \includegraphics[width=0.32\textwidth]{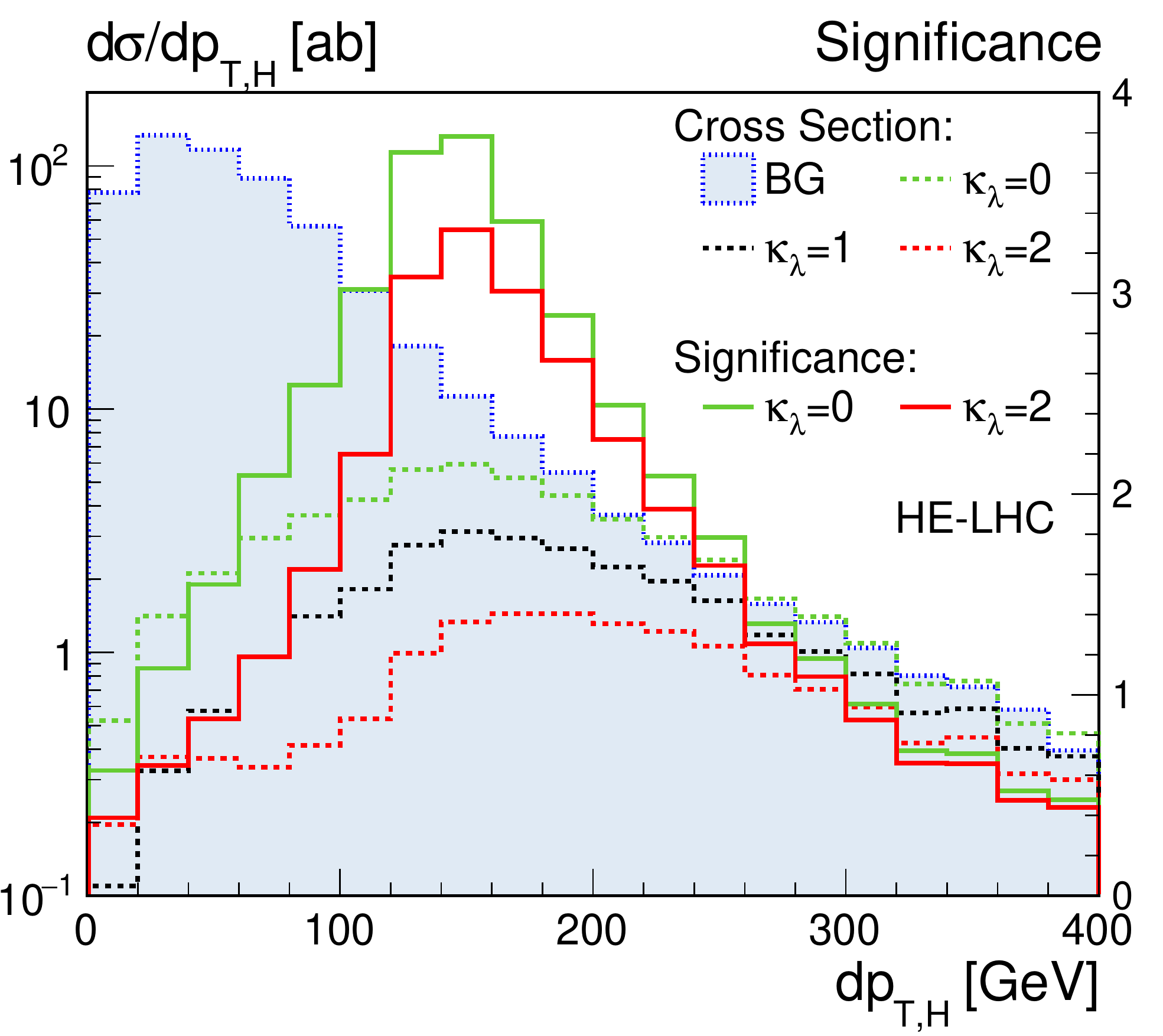}
  %\hspace*{0.02\textwidth}
  \includegraphics[width=0.32\textwidth]{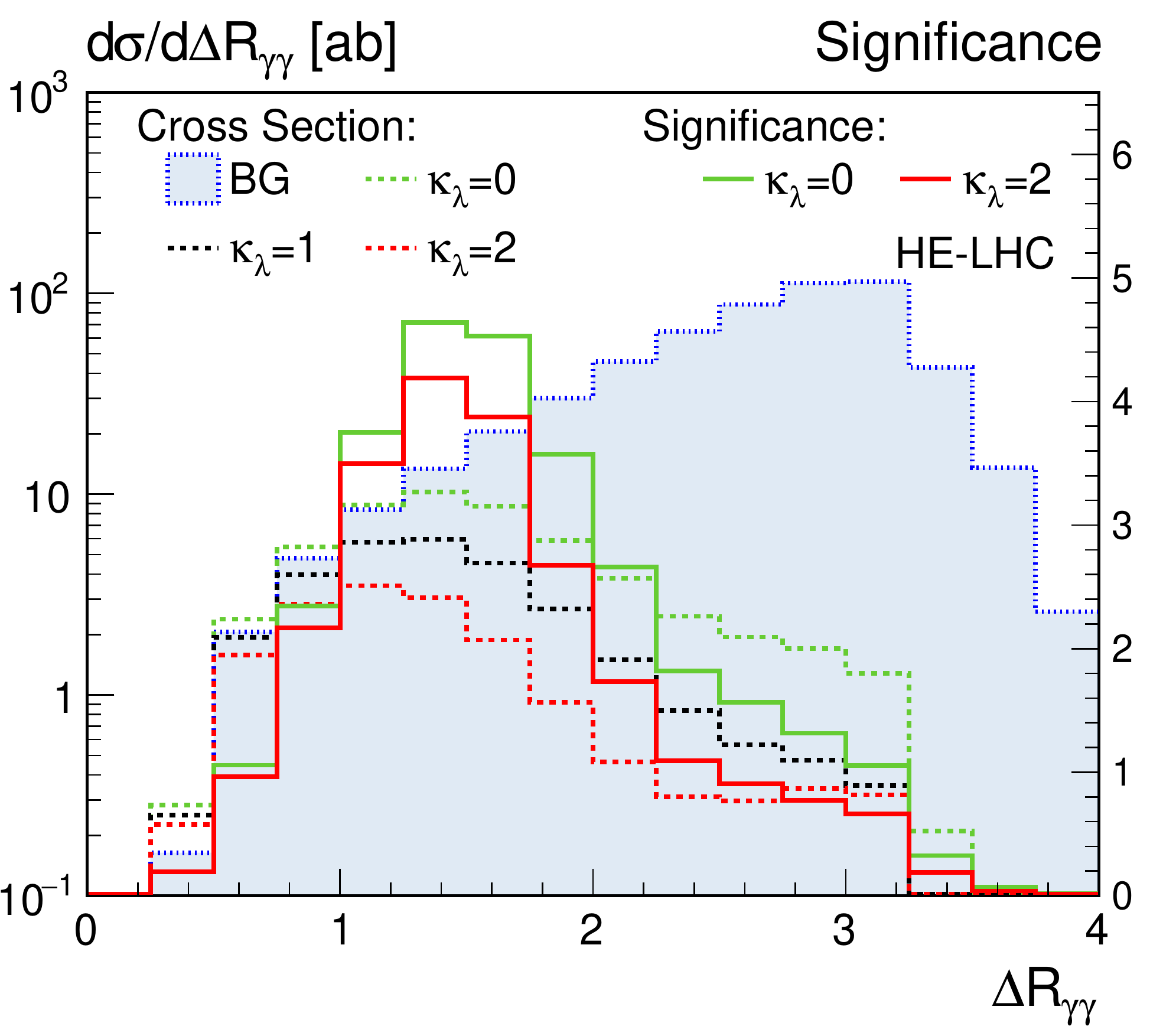} \\
  \includegraphics[width=0.32\textwidth]{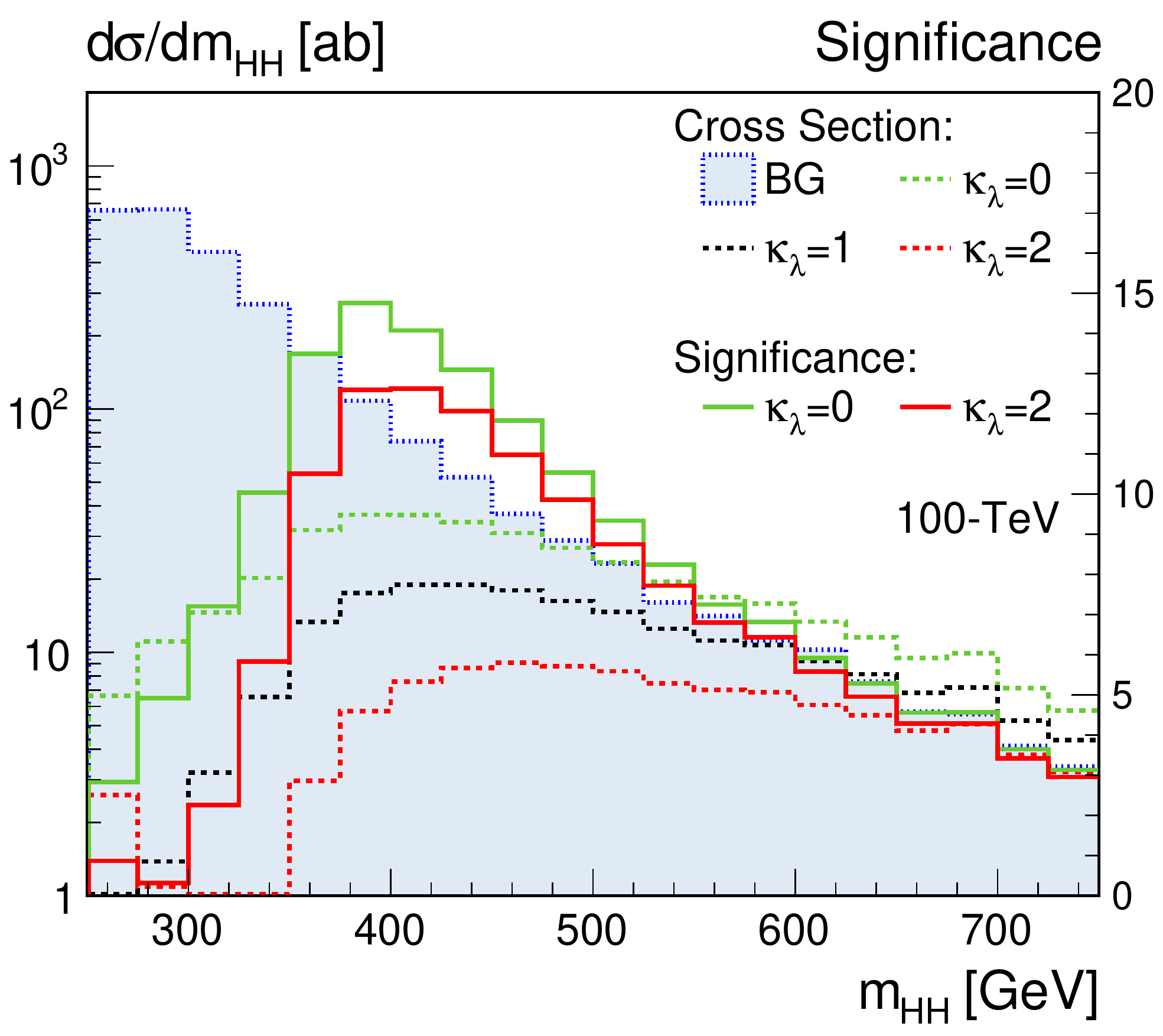}
  %\hspace*{0.02\textwidth}
  \includegraphics[width=0.32\textwidth]{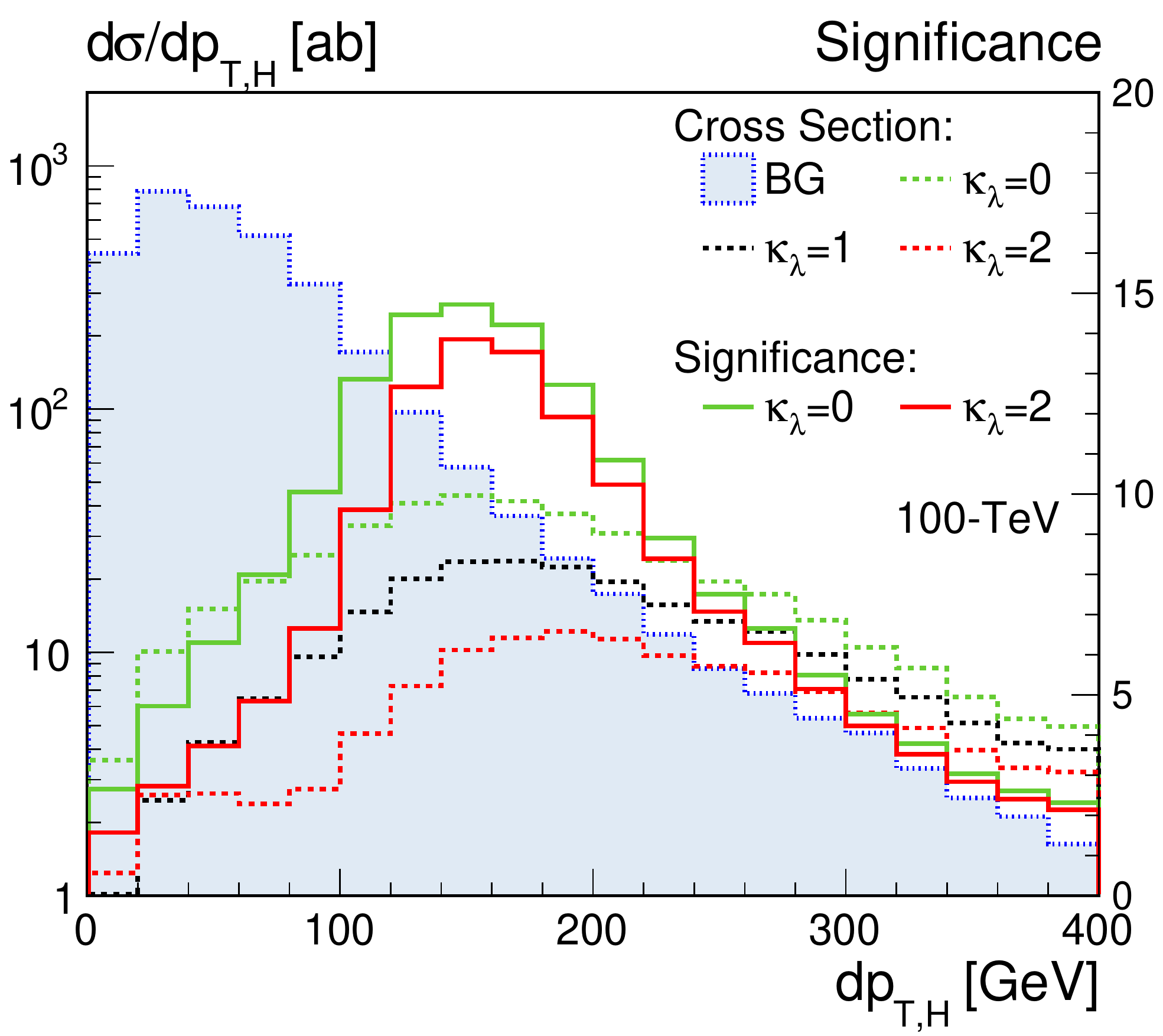}
 % \hspace*{0.02\textwidth}
  \includegraphics[width=0.32\textwidth]{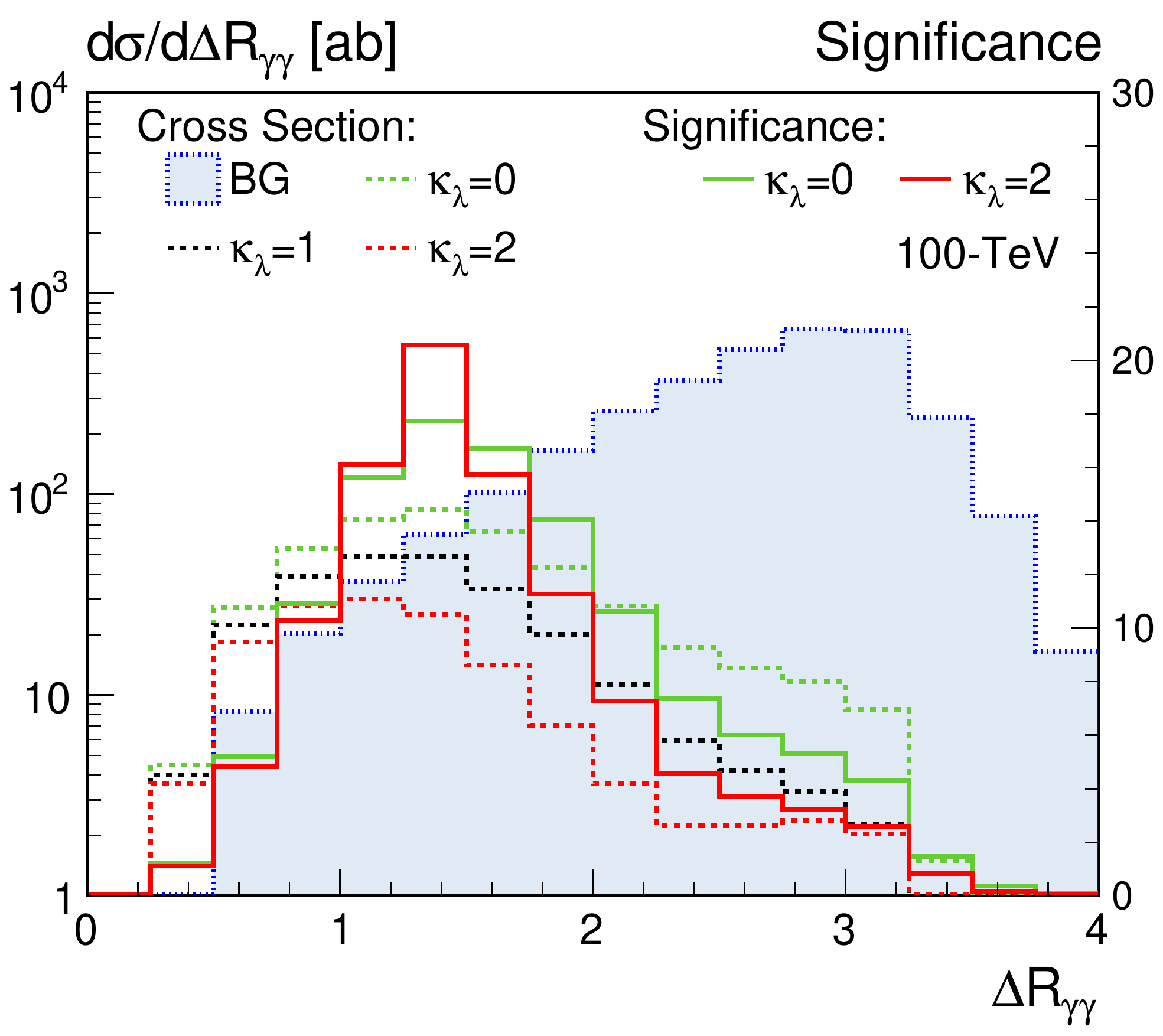}
  \caption{Kinematic distributions (dashed lines with left vertical
    axes) and significance distribution (solid lines with right
    vertical axes) assuming a Higgs self-coupling with
    $\kappa_\lambda=0,1,2$. The significance describes the
    discrimination of an anomalous self-coupling $\kappa_\lambda \neq
    1$ from the SM hypothesis $\kappa_\lambda = 1$. The
    results are for the HE-LHC (upper row) and for the 100~TeV
    collider (lower row).}
\label{fig:madmax_diff}
\end{figure*}
%---------------------------------------------

%%%%%%%%%%%%%%%%%%%%%%%%%%
\section{The Mother of Distributions}
\label{sec:features}

As depicted in Fig.~\ref{fig:feyn1}, Higgs pair production receives
contributions from a triangular loop diagram combined with the Higgs
self-coupling and from a box or continuum diagram (plus a crossing
diagram), where over most of phase space the box contribution
completely dominates the total rate.  While we can define a number of
kinematic observables describing the continuum backgrounds, the
measurement of the Higgs self-coupling relies on a simple $2 \to 2$
process with two independent kinematic variables.

Three distinct phase space regions provide valuable information on a
modified Higgs self-coupling, both from a large destructive interference between
 the triangle and box contributions. First, there is the threshold~\cite{hh-early,hh-ww}
 in the partonic center of mass
energy
\begin{align}
m_{hh}^\text{(th)} \approx 2 m_h \; .
\end{align}
In the absence of hard additional jets, the di-Higgs invariant mass is
identical to the partonic collider energy $s \equiv m_{hh}^2$.  Note
that this threshold is below $2m_t$.  Based on the effective
Higgs--gluon Lagrangian~\cite{low_energy} we can write the
corresponding amplitude for Higgs pair production as
\begin{align}
\frac{\alpha_s}{12 \pi v}
\left( \frac{\kappa_\lambda \lambda_\text{SM}}{s-m_h^2} - \frac{1}{v} \right) 
\to
\frac{\alpha_s}{12 \pi v^2}
\left( \kappa_\lambda -1 \right) \stackrel{\text{SM}}{=} 0 \; .
\label{eq:higgs_pair}
\end{align}
While the heavy-top approximation is known to give a poor description
of the signal kinematics as a whole, it does describe the threshold
dependence correctly~\cite{hh-ww}. This indicates that we can search
for a deviation of the Higgs self-coupling by looking for an
enhancement of the rate at threshold. 

Second, an enhanced sensitivity to the self-coupling appears as top
mass effect. For large positive values of $\lambda$
absorptive imaginary parts lead to a
significant dip in the combined rate at the threshold $p_{T,h} \approx
100$~GeV~\cite{hh-tautau} or equivalently~\cite{madmax-hh}
\begin{align}
m_{hh}^\text{(abs)} \approx 2 m_t \; .
\end{align}
The sharpest interference dip takes place near $\lambda\approx 2$. 
For negative values of $\lambda$ the interference becomes constructive.

Finally, the triangular and box amplitudes generally have different
scaling in the limit~\cite{hh-early,hh-tautau}
\begin{align}
m_{hh}^\text{(high)} \gg m_h, m_t \; .
\end{align}
While the triangle amplitude features an explicit suppression of either $m_h^2/m_{hh}^2$ or 
$m_t^2/m_{hh}^2$ 
at high invariant mass, the box diagrams drop more slowly towards the
high-energy regime. 

The impact of all three kinematic features can be quantified
statistically and is illustrated in detail in Fig.~5 of
Ref.~\cite{madmax-hh}. They clearly indicate that essentially the full
information on the Higgs self-coupling can be extracted through a
shape analysis of the $m_{hh}$ distribution~\cite{martin}.\medskip

The practical relevance of the different kinematic regimes has to be
estimated including the variation of the signal cross section, the
number of expected events at a given collider, and the size of the
backgrounds. There exist two similar statistical approaches to answer
this problem, the \textsc{MadMax} approach based on the Neyman-Pearson
lemma~\cite{madmax} and the \textsc{MadFisher} approach based on
information geometry~\cite{madfisher}. While the latter is especially
well-suited to estimate the reach for example of precision
measurements at the LHC, we employ the former for a simple hypothesis
test. The integrated log-likelihood ratio over the full phase space or
specific kinematic regimes allows us to estimate the maximum
significance with which any multi-variate analysis will be able to
extract a signal from backgrounds or distinguish two assumed values of
the Higgs self-coupling~\cite{madmax-hh}. Throughout maximum
likelihood analysis we limit ourselves to irreducible backgrounds and
assume that statistical uncertainties dominate over the relevant phase
space regions.  Events with soft final states typically contribute
little to the search for new particles with weak-scale masses.  The
exact choice of acceptance cuts in Eq.~\eqref{eq:base_selections} and
the modeling of $b$-tagging or photon identification efficiencies
will have a negligible effect on our results. \medskip

For our numerical analysis, we account for all backgrounds discussed in
Sec.~\ref{sec:frame}, except for the $t\bar{t}h$ channel with its
significantly different final state. As part of the detailed
background analysis in Sec.~\ref{sec:ana}, we will see that this
assumption is justified. The setup is essentially identical to 
Ref.~\cite{madmax-hh}, but now using the cuts and fake rates
given in Sec.~\ref{sec:frame}. In particular, we account for the smearing of the Higgs peak as 
leading detector effect. The invariant mass distributions are smeared 
by a Gaussian with width 1.52 GeV for the $\gamma\gamma$ 
channel~\cite{CMS:2016zjv} and 12.6 GeV for the $bb$ 
channel~\cite{Vernieri:2014wfa}. The signal rate is adjusted 
to account for the loss of signal rate through a poor description 
of the tails of the distributions~\cite{madmax-hh}. This allows us to 
restrict ourself to the two Higgs mass windows $m_{bb} = 
80~...~160$~GeV and $m_{\gamma \gamma} = 120~...~130$~GeV. 
All other detector effects are left to our actual analysis in
Sec.~\ref{sec:ana}. \medskip

In Fig.~\ref{fig:madmax_diff} we first show the signal and background
distributions for three relevant kinematic variables, $m_{hh}$,
$p_{T,h}$, and $\Delta R_{\gamma \gamma}$. The transverse momentum
distributions of the two Higgs bosons will be identical, so we can
measure them either as $p_{T,\gamma \gamma}$ or as $p_{T,bb}$. Both,
for $m_{hh}$ and $p_{T,h}$ the QCD backgrounds reside at small values,
with similar signal-to-background ratios at the HE-LHC and the 100~TeV
collider. The geometric separation of the two photons from the
continuum background has to be large to generate an invariant mass
around the Higgs mass.

%---------------------------------------------
\begin{figure}[t]
  \includegraphics[width=0.40\textwidth]{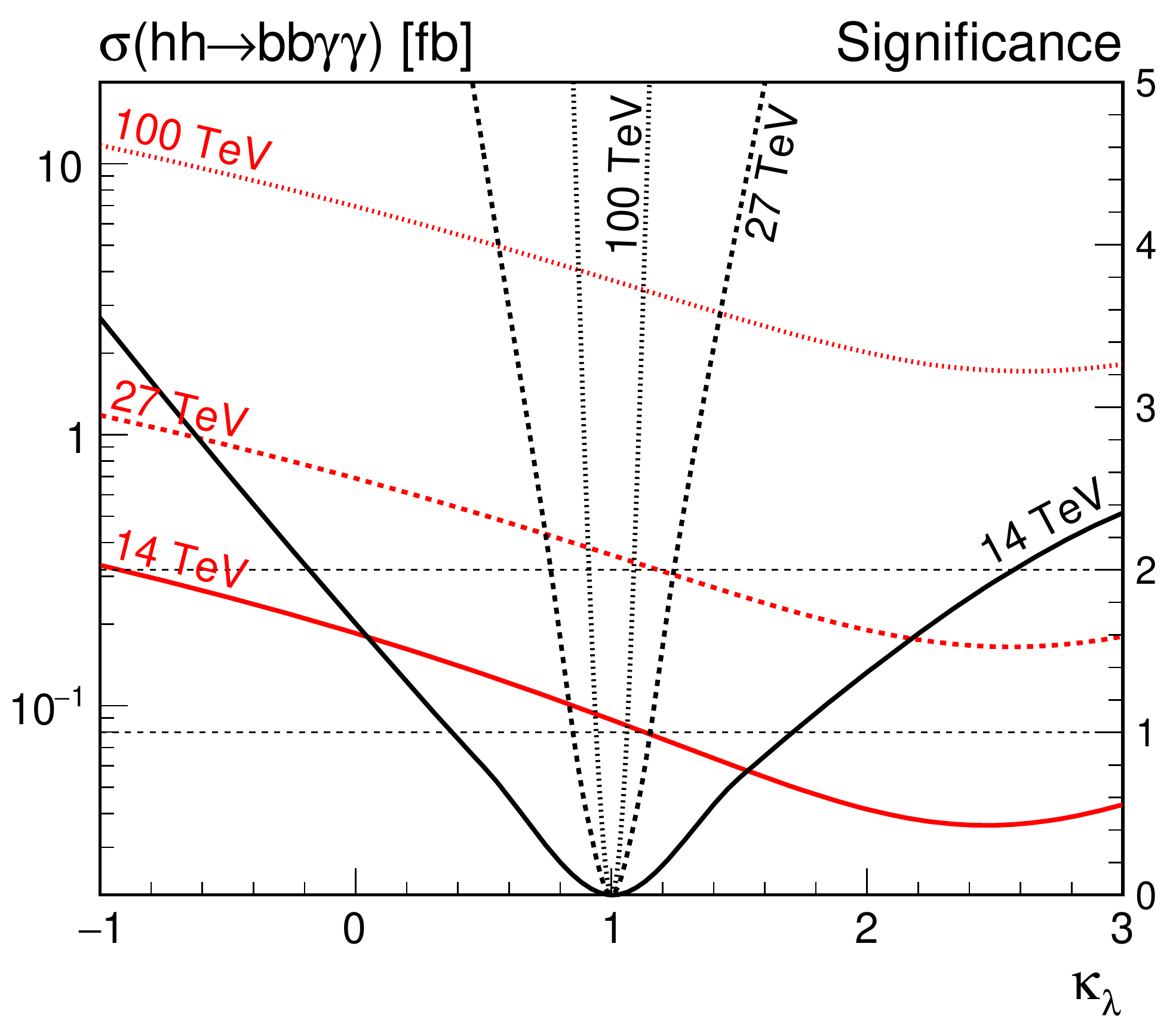}
  \caption{Higgs pair production cross section (red lines with left vertical axis)
    and maximum significance (black lines with right vertical axis) for discriminating
    an anomalous self-coupling $\kappa_\lambda \ne 1$ from the SM, as a
    function of the modified self-coupling. The results are for the
    HL-LHC, the HE-LHC, and a future 100~TeV collider,
    respectively. The HL-LHC results are taken from
    Ref.~\cite{madmax-hh}.}
\label{fig:madmax_tot}
\end{figure}
%---------------------------------------------

Also in Fig.~\ref{fig:madmax_diff}, we show how the significance of
extracting an anomalous self-coupling $\kappa_\lambda \ne 1$ depends
on these key observables.  The alternative hypothesis in this case is
the combination of the backgrounds and the signal with $\kappa_\lambda
= 1$. In addition to the signal features, the significance is limited
by the rapidly dropping backgrounds, covering both of the
above-mentioned regions with an enhanced dependence on the triangle
diagram.  In the absence of background, the significance indeed peaks
between the production threshold and the top-mass
threshold~\cite{madmax-hh}.  The drop towards large values of $m_{hh}$
is a combination of the dominance of the box diagram in the signal and
the limited number of expected signal events.  The significance with
which we can extract modified self-couplings either smaller
($\kappa_\lambda = 0$) or larger ($\kappa_\lambda = 2$) than in the
Standard Model shows a similar phase space dependence. The only
difference is a slightly harder significance distributions for
$\kappa_\lambda = 2$, an effect of the dip at
$m_{hh}^\text{(abs)}$.\medskip

Obviously, we can combine the maximum significance distributions into
a global maximum significance accumulated over the full phase
space. In Fig.~\ref{fig:madmax_tot} we show the idealized, maximum
significance with which we can hope for at the HL-LHC, the HE-LHC, and
a future 100~TeV collider. The asymmetric behavior for the HL-LHC is a
remainder of a degeneracy in the total cross section as a function of
the self-coupling, also shown in Fig.~\ref{fig:madmax_tot}.  A SM-like
rate appears when an enhanced triangle diagram overcomes the larger
box contribution and flips the sign of the amplitude. Obviously, this
degeneracy will be broken by kinematic information, for example the
$m_{hh}$ distribution. For the HE-LHC and the 100~TeV collider, the
total rate constraint becomes increasingly irrelevant for the
measurement of the self-coupling. The expected statistical error bars
are narrow and approximately symmetric around on $\kappa_\lambda = 1$. For both
future colliders, we can indeed expect a proper measurement of the
Higgs self-coupling.

%%%%%%%%%%%%%%%%%%%%%%%%%%
\section{Detector-level Analysis}
\label{sec:ana}

%-------------------------------------------------------
\begin{figure}[t]
  \includegraphics[width=.22\textwidth]{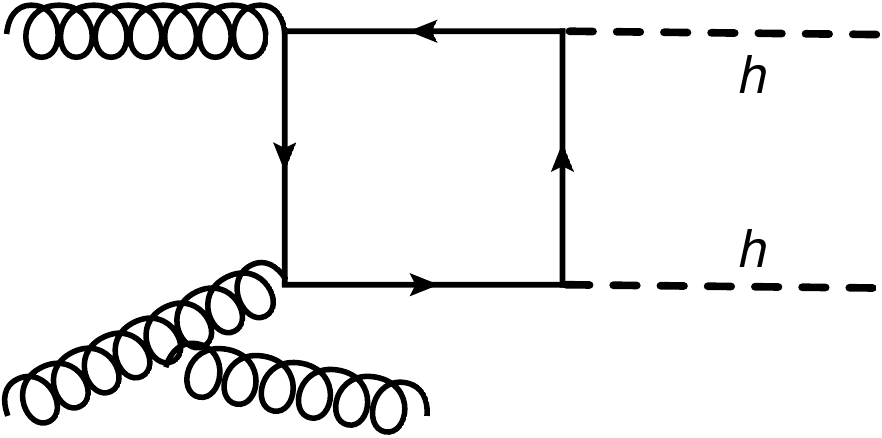}\hspace{0.2cm}
  \includegraphics[width=.22\textwidth]{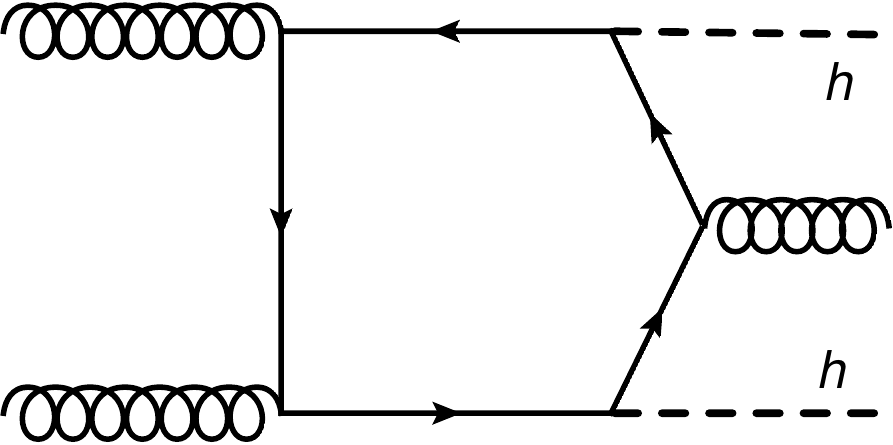}
 \caption{Representative Feynman diagrams contributing to Higgs pair production via gluon fusion including an ISR  jet at hadron colliders.}
 \label{fig:feyn2}
\end{figure}
%-------------------------------------------------------

%-------------------------------------------------------
\begin{figure*}[t]
\centering
  \includegraphics[width=.38\textwidth]{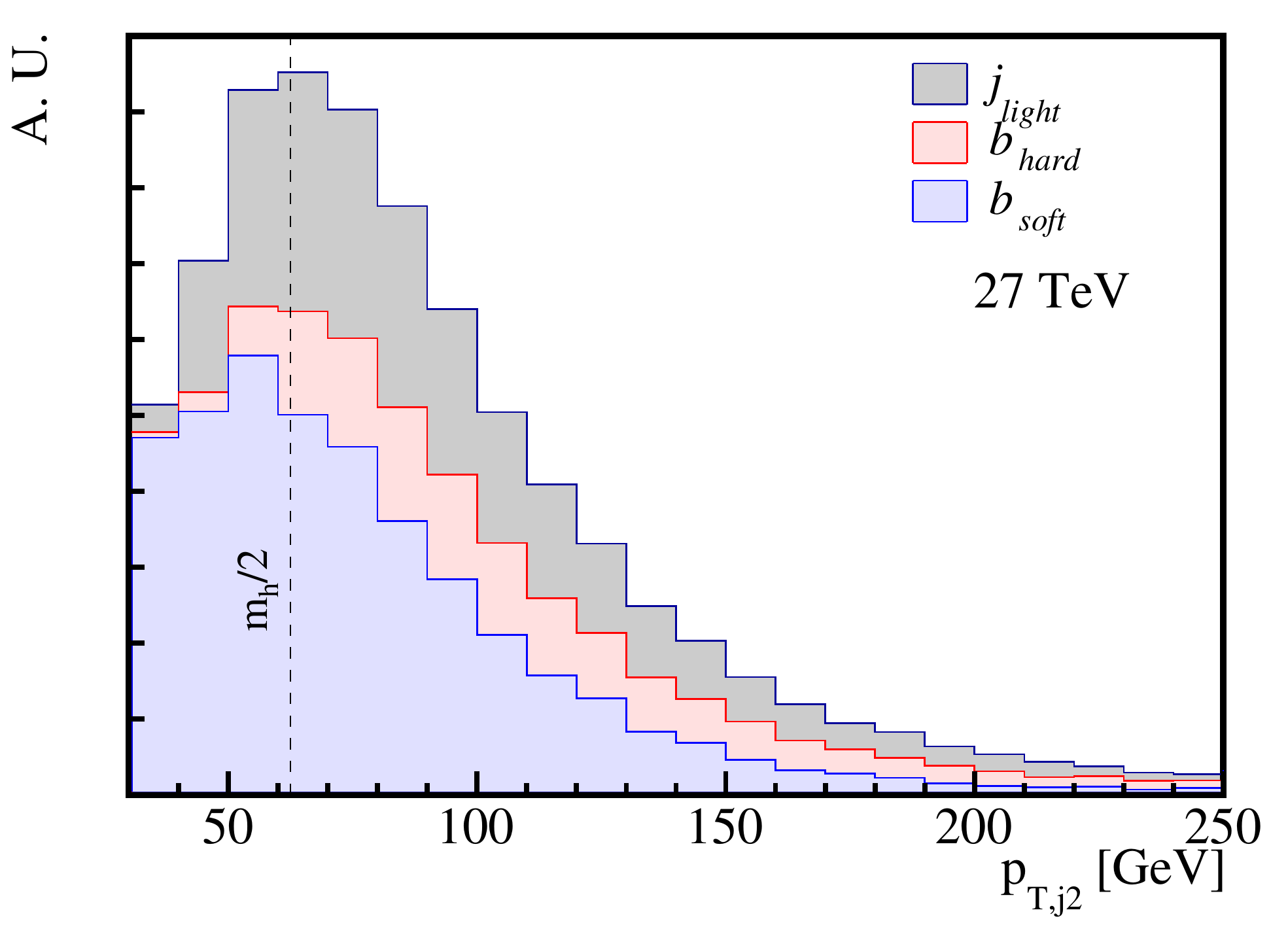}
  \hspace*{0.02\textwidth}
  \includegraphics[width=.38\textwidth]{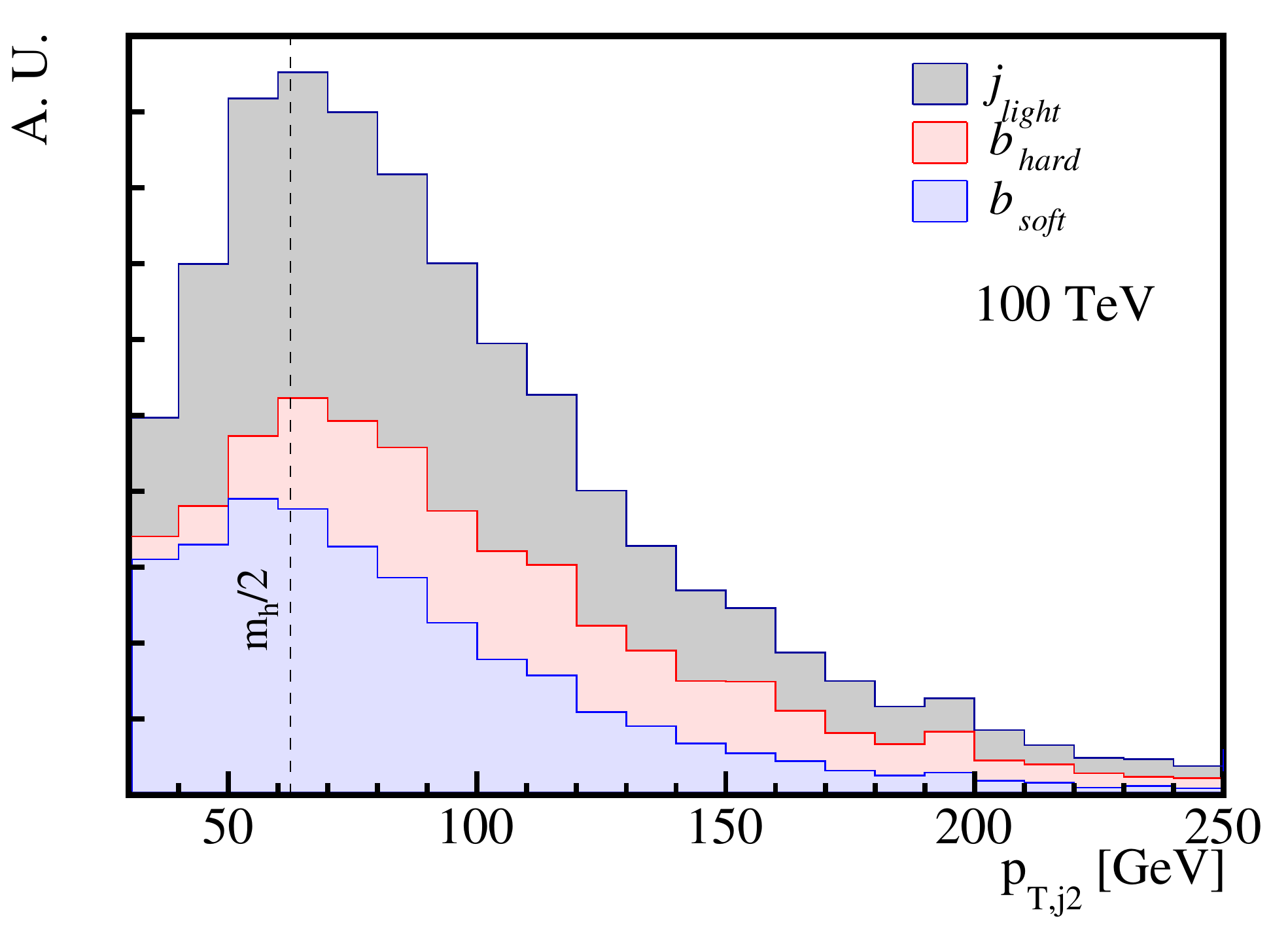}
  \vspace*{-3mm}
  \caption{Composition of the second-hardest jet in the signal sample
  after the acceptance cuts of Eq.~\eqref{eq:base_selections} for the
  HE-LHC and the 100 TeV future collider, respectively, with arbitrary units.}
 \label{fig:compo}
\end{figure*}
%-------------------------------------------------------

Following the analysis path laid out in Sec.~\ref{sec:features}, we
now design a detailed analysis strategy to extract the Higgs
self-coupling with a focus on the shape of the $m_{hh}$
distribution. Our signal is
\begin{align}
pp \to hh + X \to b\bar{b} \; \gamma \gamma + X.
\end{align}
In anticipation of increasing QCD radiation at higher energies, we 
inclusively allow extra jets in the events from initial state radiation, 
along with two tagged $b$-jets and two isolated hard photons,  
passing the acceptance cuts of Eq.~\eqref{eq:base_selections}. 

%----------------------------------------------------------
\begin{table*}[t!]
\setlength{\tabcolsep}{1pt}
\begin{tabular}{ll|rrr|rrrrr|r|cc}
\toprule
Collider & Process & \multicolumn{3}{c|}{$\kappa_{\lambda}$} & $t\bar{t}h$ & $Zh$ & $b\bar{b}\gamma\gamma$ & $jj\gamma\gamma$ & $b\bar{b}\gamma j$ & BG tot. & $S/\sqrt{S+B}_{1\iab}$ & $S/B$ 
 \\
& & 0 & 1 & 2 & & & & & & & & \\
\cmidrule{1-13}
\multirow{10}{*}{HE-LHC}
&$\sigma$ [fb]  & 0.69 & 0.36 & 0.18 & 6.43 & 0.77 & 1.24 pb & 36.6 pb & 506 pb &  &  &  
 \\
\cmidrule{2-13}
& Baseline  & 2.87K & 1.57K & 838 & 21.8K & 1.44K & 1.19M & 36M & 1.13M & 38.3M & 0.07 & $4\cdot 10^{-5}$\\
&$n_{j} \le 3$, $n_b =2$  &  648 & 356 & 190 & 954 & 389 & 200K & 67.4K & 105K &  374K & 0.15& $1 \cdot 10^{-3}$\ \\
&$\Delta m_{bb} \le 25$~GeV     & 470 & 260 & 140 & 195 & 66 & 43.7K & 10.6K  & 25.8K & 80.4K & 0.24 & 0.003 \\  
\cmidrule{2-13}
&$\Delta m_{\gamma\gamma} \le 3$~GeV  
&  459 & 253 & 136 & 197 & 63 & 1.42K & 505 & 758 & 2.94K & 1.2 & 0.09 \\
(15~ab$^{-1}$) &$\Delta m_{\gamma\gamma} \le 2$~GeV 
&  459 & 253 & 136 & 197 & 63 & 957 & 342 & 504 & 2.06K & 1.4 & 0.12 \\
&$\Delta m_{\gamma\gamma} \le 1$~GeV 
& 459 & 253 & 136 & 197 & 63 & 485 & 182 & 245 & 1.17K & 1.7 & 0.22 \\
\cmidrule{2-13}
&$\Delta m_{\gamma\gamma} \le 3$~GeV, $m_{hh}>400$ & 320 & 206 & 120 & 56 & 21 &  324 & 97 & 178 & 676 & 1.8 & 0.30 \\
&$\Delta m_{\gamma\gamma} \le 2$~GeV, $m_{hh}>400$ & 320 & 206 & 120 & 56 & 21  & 220 & 67 & 122 & 485 &2.0 & 0.42\\
&$\Delta m_{\gamma\gamma} \le 1$~GeV, $m_{hh}>400$ & 320 & 206 & 120 & 56 & 21 & 115 & 41 & 61 & 293 & 2.4& 0.70 \\
\cmidrule{1-13}
\multirow{10}{*}{100~TeV}
&$\sigma$ [fb] & 6.95 & 3.72 & 1.97 & 84.8 & 3.76 & 6.21 pb & 126 pb & 3.03 nb &  &  &   \\
\cmidrule{2-13}
& Baseline  & 51.8K & 29.8K & 16.9K & 535K & 13.1K & 13.6M & 330M & 18.6M & 363M & 0.29  & $8\cdot 10^{-5}$\cr
&$n_{j} \le 3$, $n_b =2$ & 9.22K & 5.28K & 3.02K & 18K & 2.84K & 1.79M & 773K & 1.42M  & 4.00M  & 0.48 & 0.001 \cr
&$\Delta m_{bb} \le 25$~GeV  
& 6.45K & 3.80K & 2.18K & 3.3K & 669 & 361K & 218K & 373K & 956K & 0.71 & 0.004 \\
\cmidrule{2-13}
&$\Delta m_{\gamma\gamma} \le 3$~GeV   &  6.30K &  3.70K & 2.13K & 3.12K &  653 & 8.34K & 6.06K & 8.99K & 27.2K & 3.9 & 0.14 \\
 (30~ab$^{-1}$) &$\Delta m_{\gamma\gamma} \le 2$~GeV   
&  6.30K &  3.70K & 2.13K & 3.12K &  653 & 5.66K & 4.13K &  5.99K & 19.5K & 4.4 & 0.19 \\
&$\Delta m_{\gamma\gamma} \le 1$~GeV %
&  6.30K &  3.70K & 2.13K & 3.12K &  653 &  2.82K & 1.91K & 2.99K &  11.4K & 5.5 & 0.32\\
\cmidrule{2-13}
&$\Delta m_{\gamma\gamma} \le 3$~GeV, $m_{hh}>400$                %
& 4.66K & 3.16K & 1.93K  &1.09K & 203 & 1.56K & 1.10K & 1.90K & 5.86K & 6.1 & 0.54 \\
&$\Delta m_{\gamma\gamma} \le 2$~GeV, $m_{hh}>400$               %
& 4.66K & 3.16K & 1.93K  &1.09K & 203 & 1.04K & 747 & 1.14K & 4.23K& 6.7 & 0.73 \\ 
&$\Delta m_{\gamma\gamma} \le 1$~GeV, $m_{hh}>400$              %  
& 4.66K & 3.16K & 1.93K  &1.09K & 203 &  523 & 359  &617 & 2.79K & 7.5 & 1.13 \\
\bottomrule
\end{tabular} 
\caption{Number of signal and background events for the HE-LHC and the
  100~TeV collider. We present results for $\kappa_\lambda=0,1,2$
  and the Higgs mass windows $ |m_{\gamma\gamma}-m_h|<1,2,3$~GeV. In our analysis 
  $c\bar{c} \gamma\gamma$ events are part of the $jj\gamma\gamma$ background. The
  significance is given for $1~\iab$ of data.}
\label{tab:cutflow}
\end{table*}
%----------------------------------------------------------

For the detector-level analysis we generate the signal and background samples with
\textsc{MadGraph5}+\textsc{Pythia8}~\cite{mg5,pythia8}, including one
extra jet using the \textsc{Mlm} scheme~\cite{mlm}. A representative
set of Feynman diagrams for the signal is shown in
Figs.~\ref{fig:feyn1} and~\ref{fig:feyn2}.  Higher-order corrections
are included through a next-to-leading order $K$-factor
1.6~\cite{nlo,nnlo,hh_madgraph}, neglecting possible higher-order
effects on the $m_{hh}$ distribution. We normalize the $t\bar{t}h$ and
$Zh$ to their respective NLO and NNLO rates 2.8~pb and 2.2~pb at 27~TeV
(37~pb and 11~pb at 100~TeV)~\cite{tth_zh}.  We also include the full
set of detector effects with \textsc{Delphes3}~\cite{delphes},
following the HL-LHC projections~\cite{performance}.\medskip

Jets are defined with the anti-$k_T$ algorithm  ${R=0.4}$ via
\textsc{FastJet}~\cite{fastjet}.  While the $t\bar{t}h$ background is
almost irrelevant at the 14~TeV LHC, it becomes increasingly important
at higher energies. Obviously, the more complex, high-multiplicity
final state offers many handles to tame it. We employ a simple
veto on leptons with 
\begin{align}
p_{T,\ell}>10~\gev\ \text{and}\ |\eta_\ell|<2.5 \; ,
\end{align}
combined with a veto of more than three jets passing
Eq.~\eqref{eq:base_selections}.

%-------------------------------------------------------
\begin{figure*}[t]
\centering
  \includegraphics[width=.4\textwidth]{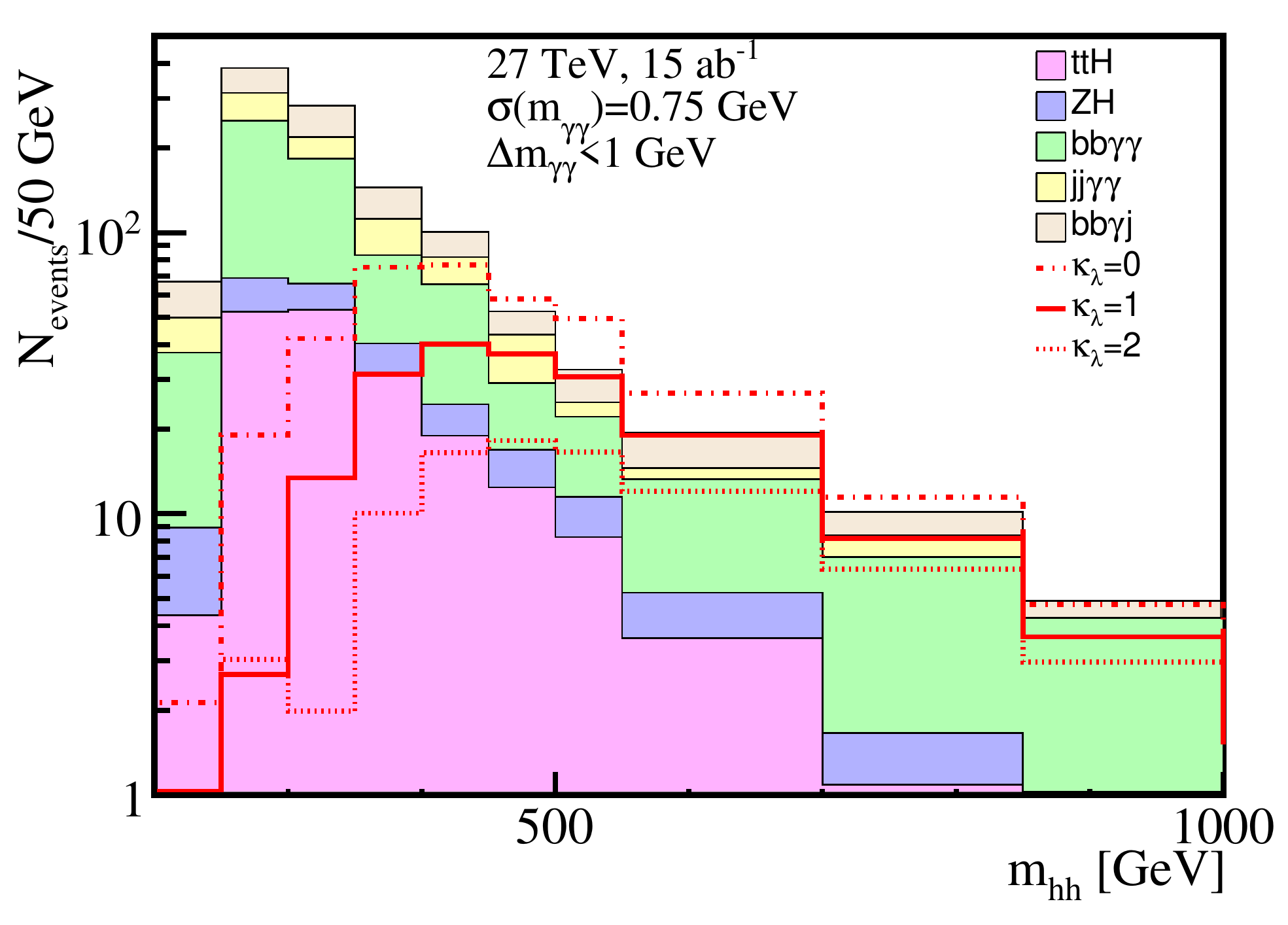}
  \hspace*{0.02\textwidth}
  \includegraphics[width=.4\textwidth]{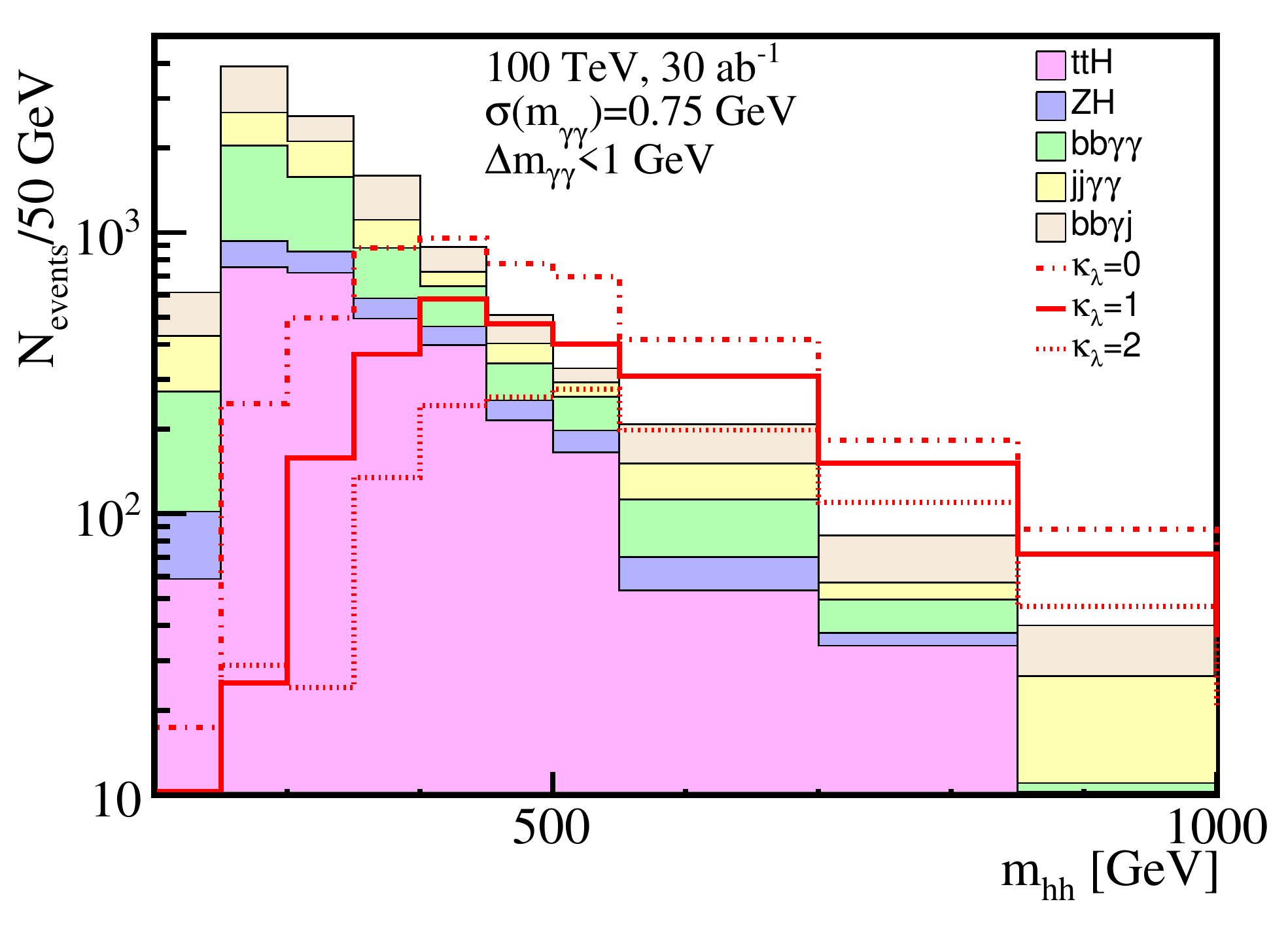}
  \vspace*{-3mm}
  \caption{Higgs pair invariant mass for the signal and backgrounds
    based on realistic simulations for the HE-LHC (left) and the 100 TeV
    future collider (right).  The $m_{\gamma\gamma}$ distribution is described
    by a Gaussian with width $0.75$~GeV.}
 \label{fig:reso}
\end{figure*}
%-------------------------------------------------------

To suppress the initially overwhelming $jj\gamma\gamma$ background, we
demand two $b$-tags among the three hardest jets. A crucial
observation is that at higher energies, initial state radiation (ISR)
often leads to a harder jet than the Higgs decay products, such that
either the hardest or second-hardest jet is not a $b$-jet for roughly
half of all events. This is illustrated in Fig.~\ref{fig:compo} as the
composition of the second-hardest parton-level jet, requiring that
both truth-level $b$-jets pass the selection of
Eq.~\eqref{eq:base_selections}. Thus, the $b$-tagging requirement
as the two leading jets should be adjusted accordingly.

Based on this observation we account for two patterns of the $p_T$
jets, $(bb,bbj)$ and $(jbb,bjb)$.  This increases our signal
efficiency by around 50\%.  Expanding this scheme to even more jets is
not effective because it eventually also increases the continuum
backgrounds and the $t\bar{t}h$ contributions. The reliability of our
Monte Carlo simulation underlying this procedure is guaranteed by the
fact that the hardest three jets are generated using multi-jet
merging.

To control the continuum backgrounds, we require two Higgs mass windows,
\begin{align}
 |m_{bb}-m_h|<25~\gev, \quad 
 |m_{\gamma\gamma}-m_h|<1~\gev  .
 \label{eq:jreslov}
\end{align}
An obvious way to enhance the Higgs pair signal is to improve the
resolution on the reconstructed photons and $b$-jets from the Higgs
decays.  We adopt the rather conservative resolution for $m_{bb}$ as
in Eq.~\eqref{eq:jreslov}. Any improvement on it in experiments would
be greatly helpful for the signal identification and background
separation.  As for the photon resolution, we illustrate this effect
by using three representative values where the $m_{\gamma\gamma}$
distribution is smeared by a Gaussian width of $0.75$, $1.5$, $2.25$~GeV,
corresponding to Higgs mass windows
\begin{align}
|m_{\gamma\gamma} - m_h| \le 1,2,3~\gev.  
 \label{eq:preslov}
\end{align}
A resolution of $1.5$~GeV has already been achieved at the
LHC~\cite{CMS:2016zjv}.\medskip

The results at this stage of the analysis are illustrated in
Table~\ref{tab:cutflow} with a full cut flow for the two collider
energies and assuming $\kappa_\lambda = 0,1,2$. We already find a
large background suppression $S/B\sim 0.09~...~0.2$ for the HE-LHC and
$0.14~...~0.3$ at a future 100~TeV collider. Requiring
${m_{hh}>400}$~GeV improves it to $S/B\sim 0.3~...~0.7$ or
$0.5~...~1.1$, respectively. This is entirely due to the rapidly
falling backgrounds as compared to the $hh$ signal,but will be at the
expense of the self-coupling determination.  The $m_{hh}$ distribution
of the signal and the different backgrounds is shown in
Fig.~\ref{fig:reso}.

%-------------------------------------------------------
\begin{figure*}[t]
\centering 
 \includegraphics[width=.4\textwidth]{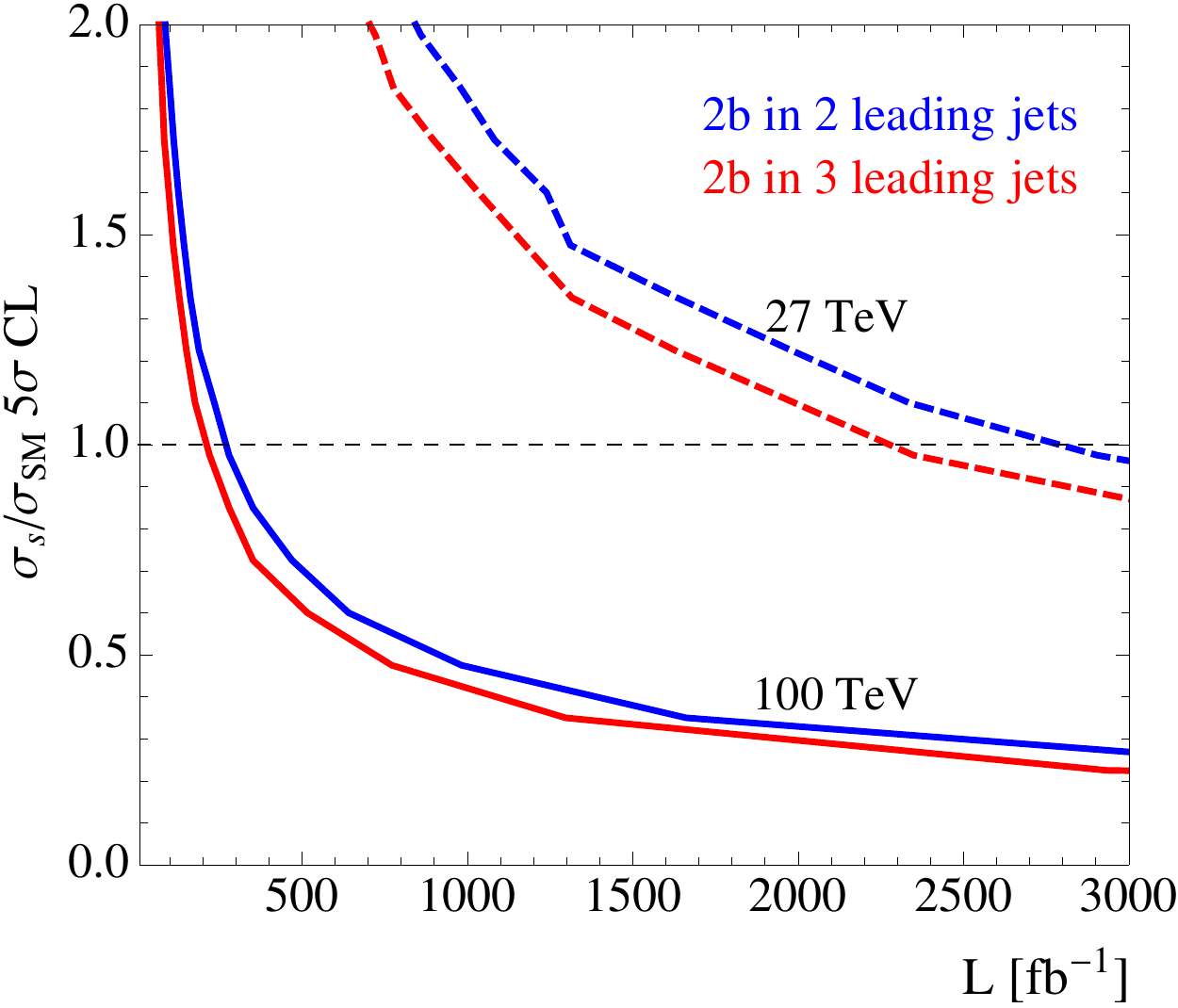} 
 \includegraphics[width=.4\textwidth]{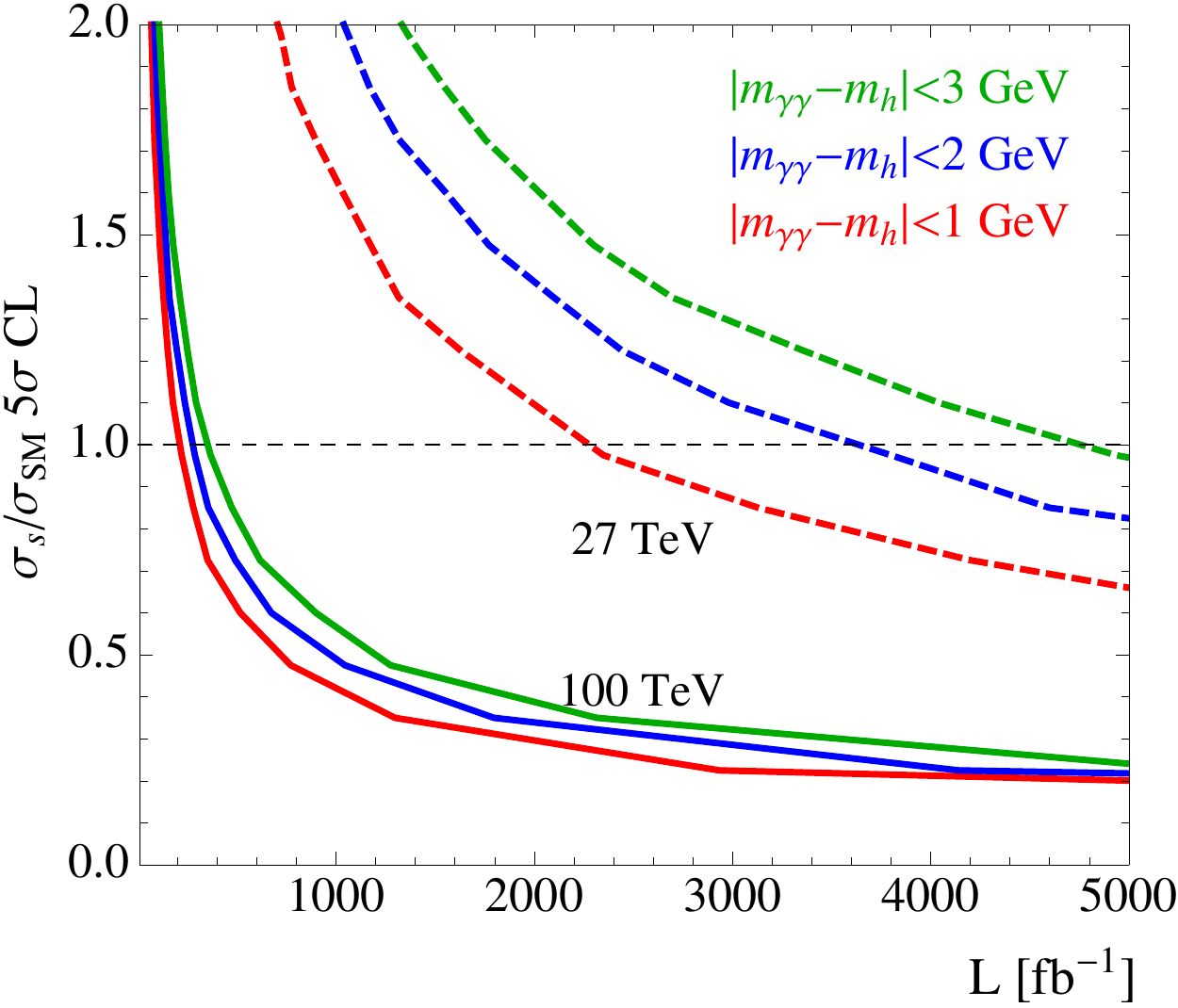}
   \caption{Luminosity required for a $5\sigma$ discover of Higgs pair
     production for the HE-LHC (dashed) and a 100~TeV collider (full).
     Left: sensitivity in terms of the total rate, demanding two
     $b$-tags among the two or three leading jets and assuming
     $|m_{\gamma\gamma}-m_h|<1$~GeV.  Right: sensitivity for three
     mass windows $|m_{\gamma\gamma}-m_h|<1,2,3$~GeV.  We assume the
     SM hypothesis with $\kappa_\lambda=1$ and use a binned
     log-likelihood analysis of the $m_{hh}$ distribution.}
 \label{fig:bound1}
\end{figure*}
%-------------------------------------------------------

The signal-to-background ratio can be strongly improved
by a better $m_{\gamma\gamma}$ resolution.  As long as most of the
$h\to \gamma\gamma$ events are captured by an
appropriate $m_{\gamma\gamma}$ window, the contributions from continuum
backgrounds can be estimated using the side-band measurements.
\medskip

Going beyond a cut-based analysis for example on $m_{hh}$, we employ a binned
log-likelihood analysis based on the CL$_{s}$ method, using the full
$m_{hh}$ distribution to extract $\kappa_{\lambda}$~\cite{read}. The
dominant backgrounds feature powerful control regions or ratio
measurements like $t\bar{t}h/t\bar{t}Z$~\cite{nimatron_yt}. Therefore,
we neglect their systematic uncertainties.  As a starting point, we
show the $5\sigma$ determination on the Higgs pair signal strength in the left
panel of Fig.~\ref{fig:bound1}, requiring two $b$-tagged jets among
the two or three leading jets.  We decompose the latter case in two 
sub-samples $(bb,bbj)$ and $(jbb,bjb)$. We see how exploring the extra-jet emission
significantly improves the significance as compared to the standard procedure
adopted in the literature.  The $5\sigma$ measurement for HE-LHC is pushed
from $2.8~\iab$ to below $2.3~\iab$. 

In the right panel of Fig.~\ref{fig:bound1} we show the discovery
reach for the Higgs pair signal as a function of the luminosity of the
HE-LHC and the 100~TeV collider. We assume three di-photon invariant
mass resolutions with three Higgs mass windows as in Eq.~\eqref{eq:preslov}
for a SM self-coupling $\kappa_{\lambda}=1$. Higgs pair production will be discovered at the
HE-LHC with approximately $2.5~...~5~\iab$ and at the 100~TeV collider
with $0.2~...~0.3~\iab$ of data, in both cases well below the design
luminosity.\medskip

\begin{figure}[t]
\centering 
  \includegraphics[width=.4\textwidth]{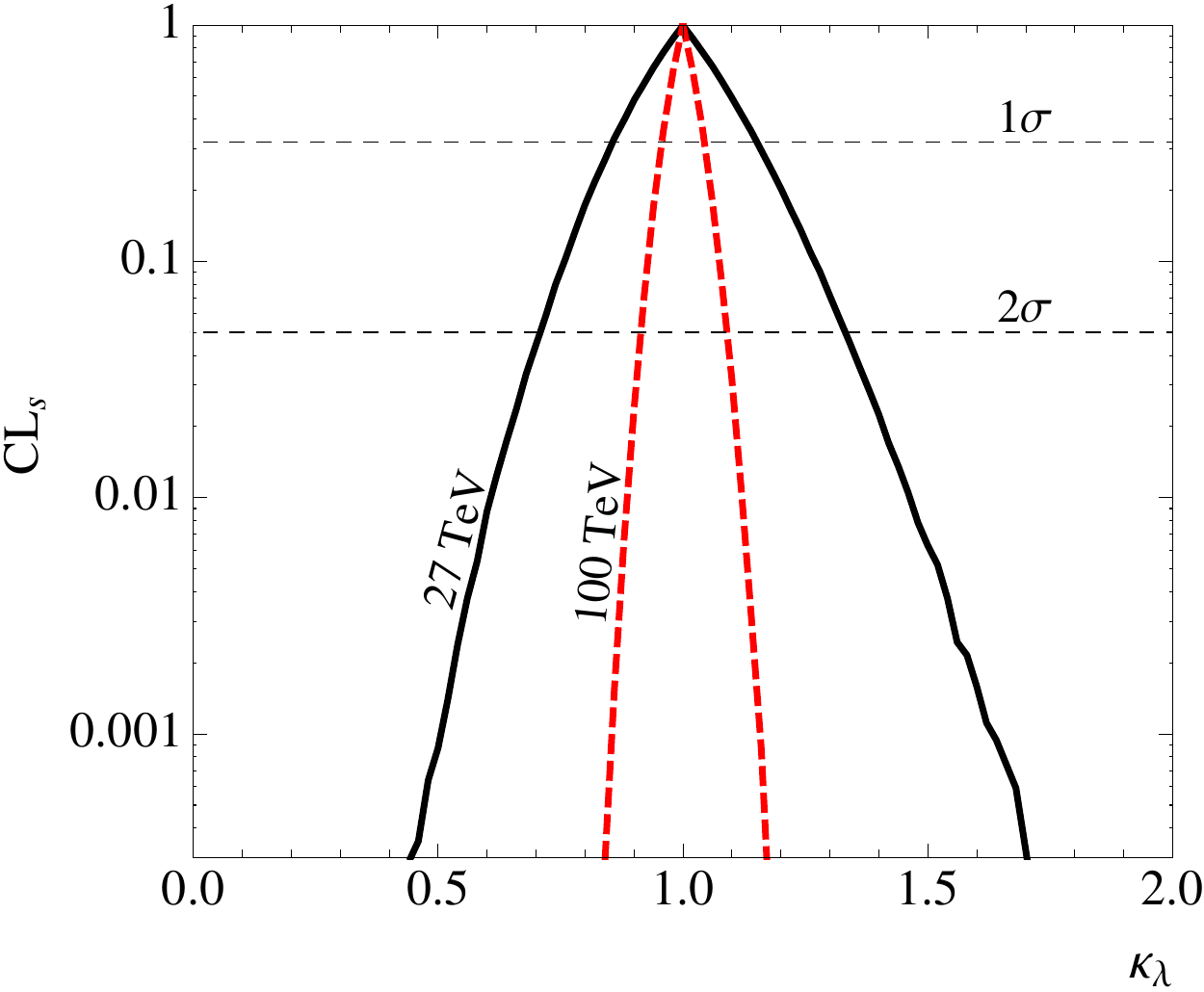}
  \caption{Confidence level for separating an anomalous Higgs
    self-coupling hypothesis from the Standard Model
    $\kappa_{\lambda}=1$.}
 \label{fig:bound2}
\end{figure}

As commented in the Introduction, there exist physics scenarios that the Higgs self-coupling could be modified at the level of order one deviation from the SM value. 
The accurate measurement of the Higgs self-coupling via Higgs pair production at future colliders has the best promise to uncover the new physics associated with the Higgs sector.
In Fig.~\ref{fig:bound2}, we show the accuracy on
this measurement. At the 68\% confidence level the triple Higgs
coupling can be measured with the precision
\begin{align}
\kappa_{\lambda} &\approx 1 \pm 15\%
\qquad \text{(HE-LHC, 27~TeV, $15~\iab$)}  , \notag \\
\kappa_{\lambda} &\approx 1 \pm 5\%
\ \,\qquad \text{(100~TeV, $30~\iab$).} 
\end{align}
At the 95\% confidence level, 
\begin{align}
\kappa_{\lambda} &\approx 1 \pm 30\%
\qquad \text{(HE-LHC, 27~TeV, $15~\iab$)}  , \notag \\
\kappa_{\lambda} &\approx 1 \pm 10\%
\qquad \text{(100~TeV, $30~\iab$).} 
\end{align}

The way to improve these expected limits towards the
mathematically-defined best reach shown in Fig.~\ref{fig:madmax_tot}
is to exploit more kinematic features and this way also suppress the
reducible $t\bar{t}h$ background.

%-------------------------------------------------------
%-------------------------------------------------------

To gain some insight on how robust our results are, we have also examined the other available choices of detector parameters, one from CMS \cite{Chatrchyan:2012jua} and the other from the CERN Yellow Report (YR)\cite{Mangano:2017tke} for the future collider (FCC). As shown in Fig.~\ref{fig:bound_appendix} in the Appendix, we find that the results are quite consistent with each other, with the YR performance being slightly better. This indicates possible room for further improvement.

%%%%%%%%%%%%%%%%%%%%%%%%%%
\section{Summary and Outlook}
\label{sec:sum}

In this paper, we have explored Higgs pair production as a direct way
to measure the Higgs self-coupling, the least-known but arguably 
the most important fundamental parameter of the Standard Model.\medskip

We first presented the production cross section for $pp \to hh$ at
future high-energy colliders in Fig.~\ref{fig:xs_hh_Ecm},
Sec.~\ref{sec:frame}. We discussed the signal rate for the process
with leading sensitivity $pp\to hh\to b\bar b\ \gamma\gamma$, and laid
out the event selection criteria in accordance with the experimental
acceptance at the LHC.

In Sec.~\ref{sec:features}, we discussed the kinematic features of the
signal and compared with the backgrounds, as shown in
Fig.~\ref{fig:madmax_diff}. The key variable is the invariant mass
distribution of the Higgs pair that presented distinctive
behaviors. We first performed a parton-level analysis that combines
the maximum significance distributions into a global maximum
significance accumulated over the full phase space, for the HL-LHC,
the HE-LHC, and a future 100~TeV collider. For both future colliders
we found excellent prospects for kinematics-based determinations of
the Higgs self-coupling as shown in Fig.~\ref{fig:madmax_tot}.\medskip

In Sec.~\ref{sec:ana}, we then carried out a search strategy based on a
rate combined with kinematic shapes with realistic simulations. The
approach is not only more powerful~\cite{hh-ww,madmax-hh}  than
a purely rate-based measurement but also
more stable against systematic and theoretical uncertainties, provided
we account for all bin-to-bin correlations.  Our method removes all
degeneracies which appear in a rate-based measurement and leads to
well-defined symmetric error bars on the modified self-coupling.

Higher energy colliders allow for including events with high
$m_{hh}$. In such more and more common configurations at high
energies, the additional jets from QCD radiation frequently surpass
the $b$-jet energy about $m_h/2$, as seen in Fig.~\ref{fig:compo}.  To
improve the signal efficiency we included at least three observable
jets, fully accounting for QCD jet radiation via the {\textsc MLM}
merging, with possibly softer $b$-jets from Higgs decays. We showed a
cut-flow in Table \ref{tab:cutflow} to illustrate the staged
improvements and to give a comparison for the two future colliders.
We further enhance our measured significances, decomposing
the samples into two sub-samples $(bb,bbj)$ and $(jbb,bjb)$.

Finally, we determined the integrated luminosity needed to reach a
$5\sigma$ significance to observe the SM $hh$ signal as shown in
Fig.~\ref{fig:bound1}.  We found that the high-energy upgrade of the
LHC to 27~TeV would reach a 5$\sigma$ observation of the Higgs 
pair production with an integrated luminosity of about 2.5 ab$^{-1}$. 
It would have the potential to reach 15\% (30\%) accuracy at 
the 68\% (95\%) confidence level to determine the SM Higgs boson 
self-coupling. A future 100 TeV collider could improve the self-coupling 
measurement to better than 5\% (10\%) at the 68\% (95\%) confidence level,
as shown in Fig.~\ref{fig:bound2}. These results roughly
agree with the optimal reach shown in Fig.~\ref{fig:madmax_tot}.
Our conclusions are quite robust 
against some moderate variations of the detector performances 
as shown in Fig.~\ref{fig:bound_appendix} in the Appendix.
In the hope of searching for effects from physics
beyond the SM, our results should provide conclusive information
weather or not the Higgs-self-interaction is modified to a level of
order one.\medskip

While our conclusions on the determination of Higgs-self-interaction
at future hadron colliders are robust and important, there is still
room for improvement. Although the final state $b\bar b\ \gamma\gamma$ is
believed to be the most sensitive channel because of the background
suppression and signal reconstruction, there exist complementary
channels such as $gg\to hh \to b\bar b\ \tau^+\tau^-$, $b\bar
b\ W^+W^-$, $b\bar b\ b\bar b$, etc. The kinematics-based measurement
and the all features related to QCD radiation at higher energies
should be equally applicable to all of them.

%%%%%%%%%%%%%%%%%%%%%%%%%%
\section{Appendix}
\label{sec:appendix}

As explained in the text, we optimize our set of selection cuts primarily to 
reduce the continuum background, which would be accompanied by large 
systematic uncertainty, and secondarily to reduce the $t\bar{t}h$ background, 
which is the largest background component with a Higgs mass peak structure.
To achieve the above optimization, we take the photon identification working point 
with a reasonably efficient jet-fake rejection~\cite{performance}, and require the
additional jet veto ($n_j \le 3$).

We believe our selection is almost optimal, but for completeness, we assess the effects 
of applying different efficiencies taken in the literature and provide the final sensitivities 
assuming those numbers. For comparison, we have worked on  two different efficiency
scenarios found for the CMS projections~\cite{Chatrchyan:2012jua}
and in the CERN Yellow Report (YR)~\cite{Mangano:2017tke} for the study of Future Circular Colliders (FCC).

%-------------------------------------------------------
\begin{figure}[t]
\centering 
  \includegraphics[width=.4\textwidth]{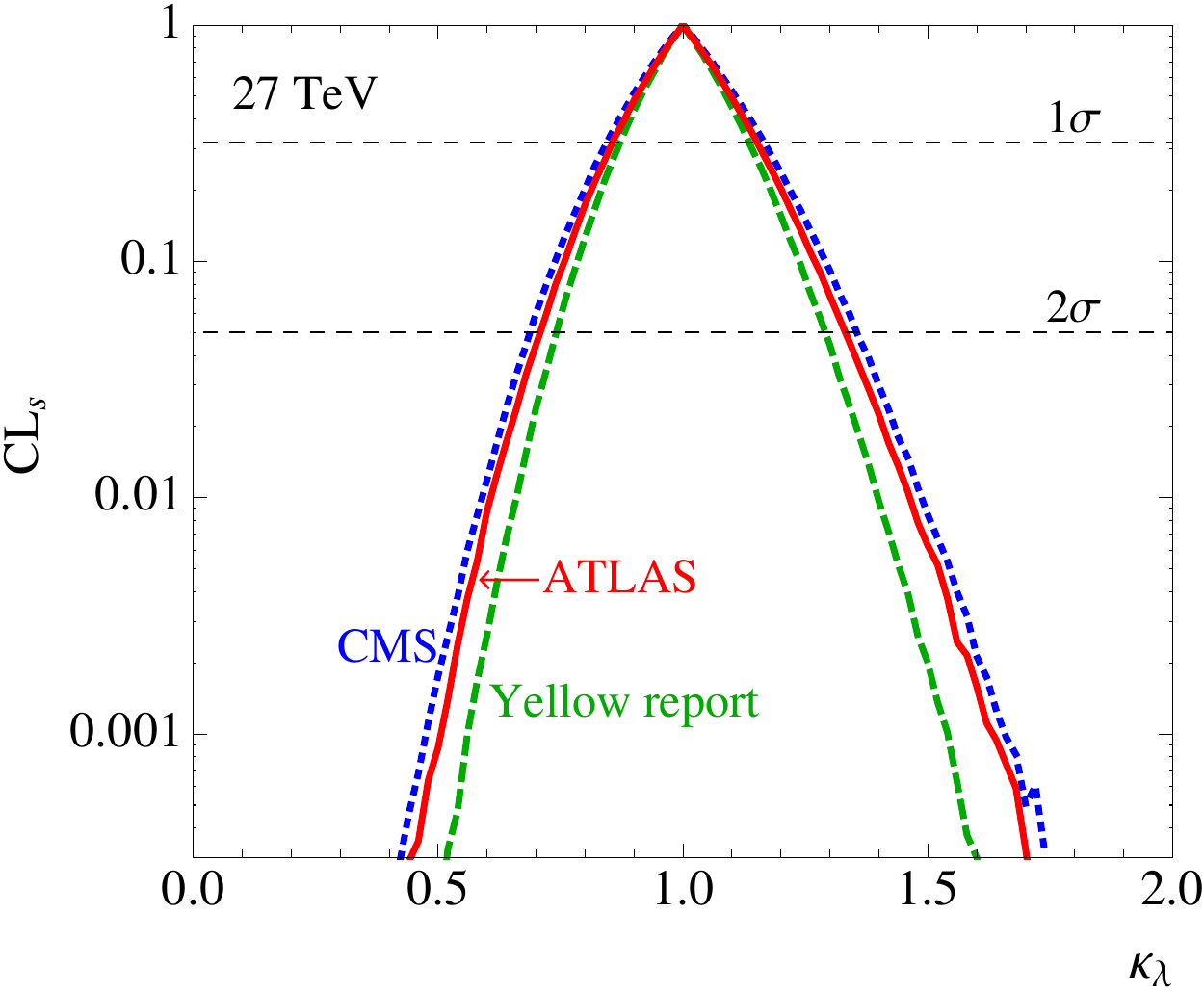}
  \includegraphics[width=.4\textwidth]{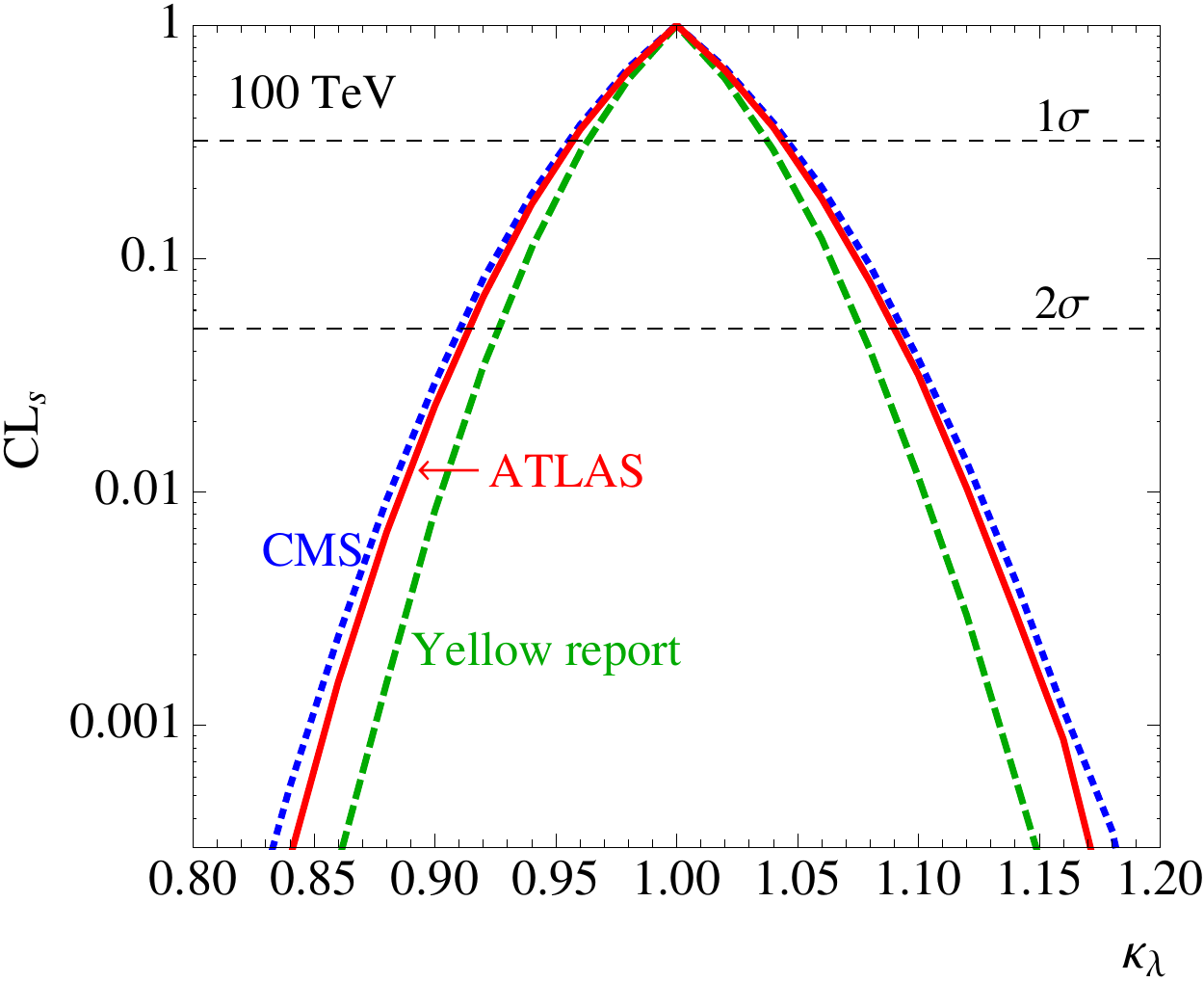}
  \caption{Comparison of the final confidence level for separating an anomalous Higgs
    self-coupling hypothesis from the Standard Model $\kappa_{\lambda}=1$ for several 
    efficiency choices. We display the results for 27~TeV (top panel) and 100~TeV (bottom panel)}
 \label{fig:bound_appendix}
\end{figure}
%-------------------------------------------------------

We adopt the fitted CMS projections as follows:
\begin{eqnarray}
&&\epsilon_{\gamma\to\gamma}=0.85,\cr
&&\epsilon_{j\to\gamma}=\begin{cases}
0.0113\exp(- \frac{p_T}{26.3~\gev})  \ \ [p_T<100~\gev]\cr
0.0025  \ \ \ \ \ \ \ \ \  \ \ \ \ \ \ \ \ \ \ \ \ \ [p_T \ge 100~\gev]   
\end{cases},\cr
&&\epsilon_{b\to b}=0.85\tanh\left(\frac{p_T}{400~\gev}\right)\frac{25.0}{1+ p_T/15.9~\gev}, \cr  
&&\epsilon_{c\to b}=0.25\tanh\left(\frac{p_T}{55.6~\gev}\right)\frac{1}{1+ p_T/769~\gev},\cr
&&\epsilon_{j\to b}=0.01.
\end{eqnarray}
The efficiency set used in the YR is the following:
\begin{eqnarray}
&&\epsilon_{\gamma\to\gamma}=0.9, \ \ \ 
\epsilon_{j\to\gamma}=0.01\exp\left(-\frac{p_T}{30~\gev}\right),\cr
&&\epsilon_{b\to b}=0.75,\ \ \ 
\epsilon_{c\to b}=0.1,\ \ \ 
\epsilon_{j\to b}=0.01.
\end{eqnarray}

Fig.~\ref{fig:bound_appendix} shows the comparison among the final results using the three 
different sets of the efficiencies for 27~TeV (top) and 100~TeV (bottom). The red lines show the final results 
assuming our adopted efficiencies (from the ATLAS HL-LHC projection study)~\cite{performance},
while the green  and the blue lines show those assuming the YR and the CMS ones, respectively.
Our analysis sensitivity is not much improved by taking the working points with a larger photon efficiency 
used by these two alternative references,  due to the corresponding worse light-jet rejection rate, which 
enhances the continuum background, especially the $bb\gamma j$ contribution. 

Note that we devise our analysis with a large $S/B$ by targeting to reduce the continuum background 
$bb\gamma j$ and leave the main background contributions from $t\bar{t}h$. In this way we achieve 
$S/B\sim 0.7$ against the corresponding numbers 0.45 (YR), and 0.4 (CMS), respectively, for the 27~TeV 
analysis. For the 100~TeV analysis, we achieve $S/B\sim 1.1$ against 0.6 (YR) and 0.5 (CMS). Thus, we can 
provide a more robust estimate against the systematic uncertainty of the continuum background. Additionally, 
it allows us to have a larger sensitivity from the lower $m_{hh}$ profile, a regime that is more background contaminated and that displays larger effects on the triple Higgs coupling.

\bigskip
\bigskip
%%%%%%%%%%%%%%%%%%%%%%%%%%
\begin{center} \textbf{Acknowledgment} \end{center}

We would like to thank Michelangelo Mangano and Michele Selvaggi for discussions. 
This work was supported in part by the U.S.~Department of Energy under
grant No.~DE-FG02- 95ER40896 and by the PITT PACC.  DG is supported in
part by the U.S.~National Science Foundation under the grant
PHY-1519175.  FK is supported by the U.S.~National Science Foundation
under the grant PHY-162063.  MT is supported in part by the JSPS
Grant-in-Aid for Scientific Research Numbers JP16H03991, JP16H02176,
17H05399, and by World Premier International Research Center
Initiative (WPI Initiative), MEXT, Japan. TP would like to thank Uli
Baur (15 years ago) and Michael Spannowsky for helpful discussions
concerning kinematic features of Higgs pair production.
  
%\newpage
%%%%%%%%%%%%%%%%%%%%%%%%%%

\end{document}

%% file: declare.tex
\AtBeginDocument{
  \heavyrulewidth=.08em
  \lightrulewidth=.05em
  \cmidrulewidth=.03em
  \belowrulesep=.65ex
  \belowbottomsep=0pt
  \aboverulesep=.4ex
  \abovetopsep=0pt
  \cmidrulesep=\doublerulesep
  \cmidrulekern=.5em
  \defaultaddspace= .5em
  \setlength{\tabcolsep}{0.5em}
}

\newcommand{\arxiv}[1]{\href{http://arxiv.org/abs/#1}{arXiv:#1}}

\newcommand\one{\leavevmode\hbox{\small1\normalsize\kern-.33em1}}

\newcommand{\gev}{{\ensuremath\rm GeV}}

\newcommand{\iab}{{\ensuremath\rm ab^{-1}}}

\def\slashchar#1{\setbox0=\hbox{$#1$}           % set a box for #1
   \dimen0=\wd0                                 % and get its size
   \setbox1=\hbox{/} \dimen1=\wd1               % get size of /
   \ifdim\dimen0>\dimen1                        % #1 is bigger
      \rlap{\hbox to \dimen0{\hfil/\hfil}}      % so center / in box
      #1                                        % and print #1
   \else                                        % / is bigger
      \rlap{\hbox to \dimen1{\hfil$#1$\hfil}}   % so center #1
      /                                         % and print /
   \fi}

\newcommand{\eg}{\textsl{e.g.}\;}

\newcommand{\be}{\begin{eqnarray*}}
\newcommand{\ee}{\end{eqnarray*}}

\newcommand{\bee}{\begin{eqnarray}}
\newcommand{\eee}{\end{eqnarray}}
\newcommand{\beeq}{\begin{equation}}
\newcommand{\eeeq}{\end{equation}}